\documentclass[]{emulateapj}

\shorttitle{Galaxy Evolution from HOD Modeling of DEEP2 and 
SDSS Galaxy Clustering}
\shortauthors{Zheng, Coil, \& Zehavi}

\def\Navg{\langle N \rangle}
\def\NavgM{\langle N(M)\rangle}
\def\Ncen{\langle N_{\rm cen} \rangle}
\def\Nsat{\langle N_{\rm sat} \rangle}
\def\NcenM{\langle N_{\rm cen}(M) \rangle}
\def\NsatM{\langle N_{\rm sat}(M) \rangle}
\def\sigM{\sigma_{\log M}}
\def\sigL{\sigma_{\log L}}
\def\meanLc{\langle L_c \rangle}
\def\meanLcM{\langle L_c(M) \rangle}
\def\Lmin{L_{\rm min}}

\def\Mmin{M_{\rm min}}
\def\Msun{M_\odot}

\def\hMsun{h^{-1}M_\odot}
\def\hLsun{h^{-2}L_\odot}
\def\erf{{\rm erf}}
\def\hMpc{h^{-1}{\rm Mpc}}

\def\Mpro{M_{\rm pro}}

\begin{document}

\title{Galaxy Evolution from Halo Occupation Distribution Modeling of 
DEEP2 and SDSS Galaxy Clustering} 
\author{
Zheng Zheng\altaffilmark{1,2}, 
Alison L. Coil\altaffilmark{2,3},
and
Idit Zehavi\altaffilmark{4}
}
\altaffiltext{1}{Institute for Advanced Study, Einstein Drive, 
                 Princeton, NJ 08540; zhengz@ias.edu.
}
\altaffiltext{2}{Hubble Fellow.}
\altaffiltext{3}{Steward Observatory, University of Arizona, 
                 Tucson, AZ 85721; acoil@as.arizona.edu.
}
\altaffiltext{4}{Department of Astronomy, Case Western Reserve University,
  10900 Euclid Avenue, Cleveland, OH 44106; izehavi@astronomy.case.edu.
}

\begin{abstract}
We model the luminosity-dependent projected two-point correlation function 
of DEEP2 and SDSS galaxies within the Halo Occupation Distribution (HOD) 
framework.  From this we infer the relationship between 
galaxy luminosity and host dark matter halo mass at $z\sim 1$ and at 
$z\sim 0$. At both epochs, there is a tight correlation between
central galaxy luminosity and halo mass, with the slope and scatter 
decreasing for larger halo masses, and the fraction of satellite 
galaxies decreasing at higher luminosity. Central $L_*$ galaxies reside 
in halos a few times more massive at $z\sim 1$ than at $z\sim 0$.  
The satellite fraction of galaxies more luminous than $L_*$ 
is $\sim$10\% at $z\sim 1$, compared to $\sim$20\% at $z\sim 0$.   
We find little evolution in the relation between the characteristic 
minimum mass of a halo hosting a central galaxy above a luminosity 
threshold and the mass scale of a halo that on average hosts one 
satellite galaxy above the same luminosity threshold, with the latter 
being 15--20 times the former. Combining these HOD results with 
theoretical predictions of the typical growth of halos, we establish 
an evolutionary connection between the galaxy populations at the two 
redshifts by linking $z\sim0$ central galaxies to $z\sim 1$ central 
galaxies that reside in their progenitor halos, which enables us to 
study the evolution of galaxies as a function of halo mass. We find 
that the stellar mass growth of galaxies depends on halo mass.
On average, the majority of the stellar mass in central galaxies 
residing in $z\sim 0$ low-mass halos ($\sim 5\times 10^{11}\hMsun$)
is the result of star formation between $z\sim 1$ and $z\sim 0$, while 
only a small fraction of the stellar mass in central galaxies of high 
mass halos ($\sim 10^{13}\hMsun$) is the result of star formation over 
this period. In addition, the mass scale of halos where the star
formation efficiency reaches a maximum is found to shift toward lower
mass with time. 
With appropriately defined galaxy samples at different redshifts,
future work can combine HOD modeling of the clustering with the
assembly history and dynamical evolution of dark matter halos.
This can lead to an understanding of the stellar mass growth due to both 
mergers and star formation as a function of host halo mass and provide 
powerful tests of galaxy formation theories.
In the appendix, we provide a brief 
discussion of systematic biases related to the assumption of ``one 
galaxy per halo'' in estimating the mass scale and number density of 
host halos from the observed clustering strength of galaxies.
\end{abstract}


\section{Introduction}

The spatial clustering of galaxies encodes useful information about their 
formation and evolution, which remain outstanding problems in modern
cosmology and astrophysics. Large redshift surveys at different epochs 
have enabled detailed studies of galaxy clustering and its evolution. 
In particular, the dependence of clustering on galaxy properties such
as luminosity and color provides fundamental constraints on galaxy 
formation theories. In this paper, we model the luminosity-dependence of 
the galaxy two-point correlation functions at $z\sim 1$ and $z\sim 0$ 
measured for galaxies in the DEEP2 Galaxy Redshift Survey \citep{Coil06} 
and the Sloan Digital Sky Survey (SDSS; \citep{Zehavi05} to investigate 
the evolution of galaxies over the last $\sim$7 billion years.

To make full use of the galaxy clustering measurements to test galaxy 
formation theories, one needs to extract and characterize the information
available in the clustering data in a convenient and informative form. 
In the cold dark matter (CDM) hierarchical structure formation paradigm, 
the evolution of galaxies is coupled to that of the dark matter halos, 
defined as roughly spherical, virialized regions with over-density about 
200 times that of the mean background density. The formation and evolution 
of dark matter halos are dominated by gravity and, with the help of 
improving computational power and $N$-body simulations, 
can be calculated accurately for any specified cosmological model.
Galaxy clustering data are particularly illuminating if one can 
relate galaxies to dark matter halos in a way that results in informative 
tests of galaxy formation theory. This can be achieved within the framework 
of the halo occupation distribution (HOD), which describes the statistical 
relation between galaxies and dark matter halos (see e.g., 
\citealt{Jing98a,Ma00,Peacock00,Seljak00,Scoccimarro01,Berlind02,Cooray02}).
The HOD characterizes this relation in terms of the probability distribution 
$P(N|M)$ that a halo of virial mass $M$ contains $N$ galaxies of a given type 
and the relative spatial and velocity distribution of galaxies and dark matter 
within halos.

Contemporary large galaxy surveys, such as the SDSS \citep{York00} and the 
Two-Degree Field Galaxy Redshift Survey (2dFGRS; \citealt{Colless01}), enable 
detailed studies of the $z\sim0$ galaxy population. HOD modeling, or the 
closely related ``conditional luminosity function'' (CLF) method
\citep{Yang03}, has been applied to interpret clustering data from
these local surveys (e.g., \citealt{Jing98b,Jing02,Bosch03,Magliocchetti03,
Zehavi05,Yang05}). Recently, much effort has also been placed on studying 
the clustering of high redshift galaxies, from $z\lesssim1$, such as the 
DEEP2 survey \citep{Davis03}, the COMBO-17 survey (Classifying Objects with 
Medium Band Observations in 17 filters; \citealt{Wolf04}), the NDWFS (NOAO 
Deep Wide-Field Survey; \citealt{Jannuzi99}), and the VVDS 
(VIMOS-VLT Deep Survey; \citealt{LeFevre05}), up to $z\sim 3$--$5$ in many 
surveys (e.g., \citealt{Adelberger05a,Adelberger05b,Lee05,Ouchi04,Ouchi05,
Allen05,Hildebrandt05,Daddi03}). HOD/CLF modeling has also been used to 
explain the clustering properties of these high redshift galaxies (e.g., 
\citealt{Bullock02,Moustakas02,Yan03,Zheng04,Lee05,Hamana05,Cooray05a,
Cooray05b,Cooray06,Conroy06,White07,Blake07}). 
Galaxy clustering data at different redshifts provide a goldmine of
information from which to deepen our understanding of galaxy evolution.  
As an example, \citet{White07} find evidence for merging or disruption of 
red galaxies in NDWFS from modeling the evolution of their clustering.
Recent studies of galaxy luminosity functions and two-point correlation 
functions at different redshifts with the CLF approach also reveal some 
interesting evolutionary trends, such as a brightening of the luminosity 
of central galaxies at a fixed halo mass toward higher redshifts (e.g., 
\citealt{Cooray05a}).

As galaxies evolve, they grow in both stellar and dark matter mass.  
Their stellar component continuously changes as the original stars age
and new stars form. Galaxies can also change their stellar contents 
and increase their mass as a result of mergers with other galaxies. 
Different feedback or pre-heating mechanisms, such as those caused by star 
formation, active galactic nuclei, or the photo-ionizing ultra-violet 
background, can also impact at different stages of a galaxy's life.  
Various evolutionary processes can change galaxy properties such as 
luminosity, color, mass and morphology. To observationally establish a 
connection between galaxy populations at different redshifts is therefore 
not trivial. Much of the work to identify progenitors or descendants of 
certain galaxy types from galaxy clustering has been based on rather simple 
models with limited constraining power, such as the object-conserving or 
the merging model \citep{Matarrese97,Moscardini98}. By contrast, the HOD 
framework converts the observed galaxy clustering data to a relation between 
galaxies and their host dark matter halos, placing galaxies in a 
cosmological context. Using HOD constraints at different redshifts derived 
from clustering data thus provides a physically motivated way to study 
galaxy evolution. 


The evolution of the HOD reflects how the relation between galaxies and 
halos changes with time, which can be used to test galaxy formation models 
in a more transparent way.  The evolution of dark matter halos can be solved 
with analytic approaches (e.g., \citealt{Taylor01,Taylor04,Benson02,Oguri04,
Bosch04,Zentner05}) and cosmological simulations (e.g., \citealt{Springel05}). 
Thus, the HOD evolution may potentially provide valuable constraints on the 
evolution of baryons in galaxies (e.g., accretion and consumption of gas, 
formation of stars) and put powerful tests on galaxy formation theories. 
Many recent studies of galaxy evolution have measured specific
galaxy properties (e.g., the star formation rate or the star formation 
history) as a function of the mass of the galaxy, where either the 
stellar mass or an estimate of the dynamical mass is used (e.g., 
\citealt{Heavens04,Jimenez05,Juneau05,Bundy05,Cimatti06,
Erb06a,Erb06b,Jimenez06,Noeske07,Noeske07b,Zheng07}).
The unique aspect of the approaches presented and envisioned in this paper
that make use of HOD modeling of galaxy clustering at different redshifts 
is that we are able to study the evolution of galaxy properties as a 
function of the host dark matter halo mass, which is a fundamental parameter 
in galaxy formation and evolution.

In this paper we perform HOD modeling of the luminosity dependence of the 
galaxy two-point correlation function for samples at two epochs, the 
$z\sim 1$ DEEP2 galaxies and the $z\sim 0$ SDSS galaxies, to study the 
evolution of the HOD and its implication for galaxy formation models. The 
structure of the paper is as follows. In \S~2 we briefly describe the DEEP2 
and SDSS galaxy samples used here. We define our cosmological model and 
introduce the HOD parameterization in \S~3. The modeling results are 
presented in \S~4. In \S~5 we make a detailed comparison between the HODs 
at the two epochs. In \S~6 we attempt to establish an evolutionary link 
between the two galaxy populations through the growth of dark matter halos. 
This allows us to infer the evolution of galaxies as a function of the host 
halo mass during the last $\sim$7 billion years. We also discuss the star 
formation efficiency at the two epochs. Finally, in \S~7 we summarize our 
results and discuss how to extend the study in this paper to a powerful, 
comprehensive program to constrain galaxy formation and evolution theories. 
For a simple estimate, instead of using a full HOD modeling, some applications 
of galaxy clustering assume one galaxy per halo. In the Appendix we use an 
HOD model to investigate the potential systematic errors introduced by 
this assumption.

\section{Galaxy Samples}

\subsection{DEEP2 Samples}
\label{subsec:deep2}

We use the projected two-point correlation functions measured by 
\cite{Coil06} for luminosity threshold galaxy samples from the DEEP2 Galaxy 
Redshift Survey to constrain the HOD at $z\sim1$.  The DEEP2 survey uses 
DEIMOS \citep{Faber03} on the 10-m Keck II telescope to
survey optical galaxies at $z\sim1$ in a comoving volume of approximately 
5$\times$10$^6$ $h^{-3}$Mpc$^3$. DEEP2 has measured redshifts for $>30,000$ 
galaxies in the redshift range $0.7<z<1.45$ to a limiting magnitude of 
$R_{\rm AB}=24.1$ over 3 deg$^2$ of the sky.  Technical details of the DEEP2 
survey can be found in \cite{Davis03} and \cite{Davis04}.  

Here we use the four nearly volume-limited samples of \citet{Coil06}, 
defined by thresholds in rest-frame Johnson $B$ absolute magnitude 
($M_B$). 
$K$-corrections are calculated as described in \cite{Willmer06}, and no 
corrections are made for luminosity evolution.  Each sample has a minimum 
redshift of $z=0.75$ and a maximum redshift of $z=1.0-1.2$, depending on 
luminosity, and includes 5000--11,000 galaxies.  The three brightest samples 
($M_B<-20.5$, $M_B<-20.0$, and $M_B<-19.5$) are volume-limited for blue
galaxies, while the faintest sample ($M_B<-19.0$) is missing some fainter 
blue galaxies at $0.95<z<1.0$. The samples are not entirely volume-limited 
for red galaxies, due to the $R_{\rm AB}$ selection of the DEEP2 survey
(for definitions of blue and red galaxies and a discussion of this selection 
effect see \citealt{Willmer06}). By assuming little evolution in 
the red galaxy fraction at $z\sim 1$ and using a volume-limited (but 
smaller) control sample at a slightly lower redshift, we estimate that, 
for each sample, about 10\% of the full population is missed because of 
the red galaxies being not volume-limited. We address the impact of 
this sample selection on the galaxy evolution we attempt to 
establish in  \S~\ref{sec:evolutionlink}.

\subsection{SDSS Samples}
\label{subsec:sdss}
 
We compare the DEEP2 clustering and modeling results to those obtained
from the SDSS at lower redshifts.  The SDSS (\citealt{York00,Stoughton02}) 
is an ongoing project that aims to map nearly one quarter of the sky in the 
northern Galactic cap, as well as a small portion of the southern Galactic 
cap, using a dedicated 2.5m telescope \citep{Gunn06}.  A drift-scanning 
mosaic CCD camera \citep{Gunn98} is used to image the sky in five 
photometric bandpasses to a limiting magnitude of $r \sim 22.5$. Objects 
are selected for spectroscopic follow-up using specific algorithms for the 
main galaxy sample \citep{Strauss02}, luminous red galaxy sample 
\citep{Eisenstein01}, and quasars \citep{Richards02}. To a good approximation, 
the main galaxy sample consists of all galaxies with Petrosian magnitude 
$r<17.77$, with a median redshift of $\sim0.1$. The construction of the 
large-scale structure samples is described in detail in \citet{Blanton05}. 
The radial selection function is derived from the sample selection criteria 
using the $K$-corrections of \cite{Blanton03a} and an improved version of the 
evolving luminosity function model of \cite{Blanton03b}. The angular 
completeness is characterized carefully for each sector (a unique region of 
overlapping spectroscopic plates) on the sky. 

We use here the clustering results presented in \citet{Zehavi05} for an SDSS 
sample containing some $200,000$ galaxies extending over $\sim 2500$ deg$^2$
on the sky. More specifically, we use the projected two-point correlation 
function measurements for volume-limited samples defined by thresholds in 
$r$-band absolute magnitude ($M_r$). These luminosity threshold samples span 
a redshift range of $0.02<z<0.22$.  \citet{Zehavi05} have also performed 
detailed HOD modeling of these clustering measurements. We build on these 
results, although in order to facilitate a more direct comparison to the 
DEEP2 modeling presented here, we repeat the SDSS HOD modeling with a 
slightly different and more flexible suite of models used in this work. 

The SDSS and the DEEP2 surveys have different sample selections, with
SDSS galaxies selected in rest-frame $r$-band and DEEP2 galaxies in rest-frame
$B$-band. The differences in the selection complicate the comparison 
of the HOD modeling results, a point we address throughout the paper.

\section{Cosmology and HOD Models}

\subsection{Cosmology and Halo Properties}
\label{sec:cosmology}

We adopt a spatially-flat CDM cosmological model with
Gaussian initial density fluctuations that is consistent with the 
determination from the cosmic microwave background, Type Ia supernovae, and 
galaxy clustering 
\citep[e.g.,][]{Spergel03,Spergel07,Riess04,Tegmark04,Abazajian05}.
Density parameters are assumed to be ($\Omega_m$, $\Omega_\Lambda$, 
$\Omega_b$)=(0.3, 0.7, 0.047), where $\Omega_m$, $\Omega_\Lambda$, and 
$\Omega_b$ are density parameters of the matter (CDM+baryons), dark energy, 
and baryons, respectively. The mass fluctuation power spectrum is the 
primordial power spectrum, with the spectral index assumed to be $n_s$=1, 
modified by the transfer function, which is computed using the formula given 
by \citet{Eisenstein98} with the effect of baryons taken into account. It is 
normalized such that the rms fluctuation of the linear density 
in spheres of radius 8$\hMpc$ at $z$=0 is $\sigma_8$=0.9. The Hubble 
constant is $H_0\equiv100h=70\,{\rm km\,s}^{-1}{\rm Mpc}^{-1}$.

We define a dark matter halo as an object with mean density of 200 times 
that of the background universe. With the assumed cosmological model, 
properties of dark matter halos are calculated based on numerically tested 
fitting formulas. The halo mass function is computed according to the 
formula given by \citet{Jenkins01}. For the large-scale halo bias factor, 
we adopt the formula in \citet{Tinker05}, which is a modification of that 
given by \citet{Sheth01} and is accurate for a large range of cosmological 
parameters. The density distribution of a dark matter halo of mass $M$ 
is assumed to follow the Navarro-Frenk-White (NFW) profile 
\citep{Navarro95,Navarro96,Navarro97} characterized by the concentration 
parameter $c(M)$. For $c(M)$, we use the relation given by \citet{Bullock01}, 
modified to be consistent with our halo definition,
\begin{equation}
c(M)=\frac{c_0}{1+z}(M/M_*)^\beta,
\end{equation}
where $c_0=11$, $\beta=-0.13$, and $M_*$ is the nonlinear mass at $z=0$
($M_*=7.77\times 10^{12}\hMsun$ for the adopted cosmology).

\subsection{HOD Parameterization}
\label{sec:hod}

It has proven to be useful to parameterize the HOD by separating 
contributions from central and satellite galaxies, through studies of 
subhalo HODs in high-resolution, dissipationless $N$--body simulations 
\citep{Kravtsov04} and galaxy HODs in semi-analytic galaxy formation 
model (SA) and in smoothed particle hydrodynamics (SPH) simulations 
\citep{Zheng05}. 
For galaxy samples defined by lower luminosity thresholds, a simple 
parameterization of the mean occupation function has three parameters 
\citep[e.g.,][]{Kravtsov04,Zheng05,Zehavi05} --- the mean occupation 
function $\NcenM$ of central galaxies can be represented by a step-like 
function with a characteristic minimum halo mass $\Mmin$, and the mean 
occupation function $\NsatM$ of satellite galaxies is approximated 
by a power law with the amplitude and slope as two free parameters. 
This simple three-parameter model can capture the basic features in the mean 
occupation function predicted by galaxy formation models. 

However, there are some advantages in choosing a parametrization more 
flexible than the simple one. For example, 
at lower halo mass, the mean occupation function of satellites predicted 
by galaxy formation models drops faster than the extrapolation of the high
mass power law \citep{Kravtsov04,Zheng05}. Since, in the low-mass range, 
satellites have a substantial contribution to the galaxy two-point 
correlation function, with the simple three-parameter model, the best-fit 
power law slope in the satellite mean occupation function tends to reflect 
the effective slope in the low-mass range \citep[e.g.,][]{Zehavi05}. That 
is, the overall slope is adjusted to fit the amplitude of $\Nsat$ in the 
low-mass range, and it may become a poor description for the slope at high 
halo mass.

In this paper we adopt a slightly more flexible parameterization with
five parameters, motivated by the results presented in \citet{Zheng05}.
The mean occupation function of the central galaxies is a step-like function 
with a soft cutoff profile to account for the scatter between galaxy 
luminosity and host halo mass. That of the satellite galaxies is a power law
modified by a low-mass cutoff profile in better agreement with predictions 
of galaxy formation models. This more flexible parameterization includes 
two parameters for the mean occupation function of central galaxies and 
three for that of the satellite galaxies, which are described in detail below.

The mean occupation function of central galaxies is a step-like function
parameterized by 
\begin{equation}
\label{eqn:Ncen}
\NcenM=\frac{1}{2} 
\left[1+\erf\left(\frac{\log M-\log\Mmin}{\sigM}\right)\right],
\end{equation}
where $\erf$ is the error function 
\begin{equation}
\label{eqn:erf}
\erf(x)=\frac{2}{\sqrt{\pi}} \int_0^{\rm{x}} e^{-t^2} dt .
\end{equation}
There are two free parameters: $\Mmin$, the characteristic minimum mass of 
halos that can host central galaxies above the luminosity threshold, 
and $\sigM$, the width of the cutoff profile.  Note that $\Mmin$ here can 
be interpreted as the mass of such halos for which
half of them host galaxies above the given luminosity threshold, i.e., 
$\langle N_{\rm cen}(M_{\rm min})\rangle=0.5$. It is not identical to those 
used in three-parameter models (e.g., \citealt{Zehavi05}): 
$\langle N_{\rm cen}(M_{\rm min})\rangle=1/e$ for an exponential 
cutoff profile and $\langle N_{\rm cen}\rangle$ changes from 0 to 1 at 
$\Mmin$ for a sharp cutoff profile. Nevertheless, in all cases, $\Mmin$
characterizes the minimum mass scales of the host halos.
In the three-parameter model, $\NcenM$, which uses one parameter $\Mmin$, 
can also have a soft cutoff profile, e.g., an exponential profile as in 
\citet{Zehavi05}. However, the two parameters $\Mmin$ and $\sigM$ of the 
cutoff profile in equation~(\ref{eqn:Ncen}) have a more physical meaning. 
To see this, we briefly review the derivation of the form of $\NcenM$. The 
motivation for such a form is the relation between the central galaxy 
luminosity and the host halo mass predicted by galaxy formation models. 
Based on predictions of SA and SPH galaxy formation models, \citet{Zheng05} 
show that the distribution of the central galaxy luminosity $L_c$ at fixed 
halo mass $M$ can be described by a log-normal distribution, and here we 
write it as 
\begin{equation}
\label{eqn:PLc}
P(\log L_c|M) = \frac{1}{\sqrt{2\pi}\sigL}
      \exp\left[-\frac{(\log L_c - \log \meanLcM)^2}{2\sigL^2}\right].
\end{equation}
In halos of mass $M$, the mean occupation function of the central galaxies 
above a luminosity threshold $\Lmin$ is an integration of 
equation~(\ref{eqn:PLc}) over $\log L_c$, and the result turns out to have the 
same form as equation~(\ref{eqn:Ncen}) but with the argument of the 
$\erf$ function being $[\log \Lmin - \log \meanLcM]/(\sqrt{2}\sigL)$. If 
the mass range of the cutoff profile is not large so that we can approximate 
the mean luminosity $\meanLcM$ of central galaxies in halos of mass $M$ as 
$\meanLcM \propto M^p$, it is straightforward to see the meaning of the 
parameters in equation~(\ref{eqn:Ncen}) ---
$\Mmin$ is the mass of halos in which the mean luminosity of central galaxies
is the luminosity threshold $\Lmin$, and the width of the cutoff profile 
is related to the scatter of the central galaxy luminosity in halos of
mass $M$ as $\sigM=\sqrt{2}\sigL/p$. Therefore, by studying the HODs of 
galaxy samples with different luminosity thresholds, we would learn the 
distribution (mean and scatter) of central galaxy luminosity as a function
of halo mass, from the cutoff profile and the mass scale of $\NcenM$. We
follow the above explanation when interpreting our modeling results.

According to predictions from $N$--body, SA, and SPH galaxy formation 
models \citep{Kravtsov04,Zheng05}, the mean 
occupation function of satellite galaxies $\NsatM$ approximately follows 
a power law at the high halo mass end with the slope close to unity.
At lower mass, $\NsatM$ drops steeper than the power-law extrapolation
and can be parameterized by $[(M-M_0)/M_1^\prime]^\alpha$, where $M_0$
is the mass scale of the drop, $M_1^\prime$ characterizes the amplitude,
and $\alpha$ is the asymptotic slope at high halo mass. 
Applying the same cutoff profile of the central galaxies, we assume the 
following form for the mean occupation of satellite galaxies,
for $M>M_0$: 
\begin{equation}
\label{eqn:Nsat}
\NsatM=\frac{1}{2} 
\left[1+\erf\left(\frac{\log M-\log\Mmin}{\sigM}\right)\right]
\left(\frac{M-M_0}{M_1^\prime}\right)^\alpha.
\end{equation}

For modeling the two-point correlation function of galaxies, we also need to 
know the second moment of the occupation number in addition to the mean. 
Central galaxies simply follow the nearest-integer distribution 
\citep{Berlind02,Zheng05}, and for satellite galaxies, we assume a Poisson 
distribution that is consistent with theoretical predictions 
\citep{Kravtsov04,Zheng05}. The spatial distribution of galaxies inside 
halos is assumed to be the same as the dark matter that follows the NFW
profile, a reasonable assumption on scales we model 
\citep[e.g.,][]{Nagai05,Maccio06}. 
Overall, our basic HOD parameterization has a total
of five parameters --- two for $\NcenM$ ($\Mmin$ and $\sigM$) and three for 
$\NsatM$ ($M_0$, $M_1^\prime$, and $\alpha$). With respect to the
simple three-parameter model used in \citet{Zehavi05}, our five-parameter model
essentially introduces one additional parameter to $\NcenM$ and one to 
$\NsatM$ to characterize the cutoff profiles at low halo mass. 

For theoretical calculations of galaxy two-point correlation functions, we 
adopt the method proposed in \citet{Tinker05}, which improves that 
presented in \citet{Zheng04} by incorporating a more accurate treatment of 
the halo exclusion effect. The method is calibrated and tested using mock 
catalogs and can reach an accuracy of 10\% or better in calculating galaxy 
two-point correlation functions. More specifically, we adopt the 
``$\bar{n}_g^\prime$--matched'' approximation (see their Appendix B) for 
a more efficient calculation.

\section{HOD Modeling Results for DEEP2 and SDSS Galaxies}

\begin{deluxetable*}{cccrrcccrr}
\tablewidth{0pt}
\tablecolumns{8}
\tablecaption{\label{tab:hodfit}
Best-fit HOD Parameters for DEEP2 and SDSS Galaxy Samples\tablenotemark{a}}
\tablehead{Sample  & $\log\Mmin$ & $\sigM$ & $\log M_0$ & $\log M_1^\prime$ & $\alpha$ & $\log M_1$\tablenotemark{b} & $b_g$\tablenotemark{b} & N\tablenotemark{c} & $\chi^2$}
\startdata
\cutinhead{DEEP2}
$M_B<-19.0$ & 11.64$_{-0.08}^{+0.08}$ &  0.31$_{-0.19}^{+0.19}$ 
            & 12.02$_{-0.86}^{+0.93}$ & 12.57$_{-0.47}^{+0.44}$ 
            &  0.89$_{-0.34}^{+0.23}$ & 13.00$_{-0.10}^{+0.10}$
            &  1.22$_{-0.01}^{+0.01}$ & 18 &  8.4 \cr
$M_B<-19.5$ & 11.83$_{-0.07}^{+0.08}$ &  0.30$_{-0.19}^{+0.18}$ 
            & 11.53$_{-0.65}^{+0.65}$ & 13.02$_{-0.10}^{+0.11}$ 
            &  0.97$_{-0.11}^{+0.11}$ & 13.06$_{-0.08}^{+0.08}$
            &  1.32$_{-0.01}^{+0.01}$ & 18 &  4.9 \cr
$M_B<-20.0$ & 12.07$_{-0.09}^{+0.09}$ &  0.37$_{-0.19}^{+0.17}$ 
            &  9.32$_{-0.96}^{+1.53}$ & 13.27$_{-0.06}^{+0.06}$ 
            &  1.08$_{-0.06}^{+0.05}$ & 13.28$_{-0.06}^{+0.06}$
            &  1.44$_{-0.01}^{+0.01}$ & 18 &  7.9 \cr
$M_B<-20.5$ & 12.63$_{-0.11}^{+0.11}$ &  0.82$_{-0.09}^{+0.09}$ 
            &  8.58$_{-0.94}^{+0.97}$ & 13.56$_{-0.06}^{+0.06}$ 
            &  1.27$_{-0.12}^{+0.12}$ & 13.58$_{-0.06}^{+0.06}$
            &  1.47$_{-0.02}^{+0.02}$ & 18 & 14.8 \cr
\cutinhead{SDSS}
$M_r<-18.0$ & 11.35$_{-0.07}^{+0.07}$ &  0.25$_{-0.16}^{+0.18}$ 
            & 11.20$_{-0.54}^{+0.66}$ & 12.40$_{-0.12}^{+0.15}$ 
            &  0.83$_{-0.07}^{+0.09}$ & 12.47$_{-0.10}^{+0.10}$ 
            &  0.91$_{-0.02}^{+0.02}$ & 12 &  12.7 \cr
$M_r<-18.5$ & 11.46$_{-0.06}^{+0.06}$ &  0.24$_{-0.16}^{+0.18}$ 
            & 10.59$_{-0.82}^{+0.86}$ & 12.68$_{-0.08}^{+0.09}$ 
            &  0.97$_{-0.05}^{+0.06}$ & 12.70$_{-0.08}^{+0.08}$ 
            &  0.95$_{-0.02}^{+0.02}$ & 12 &  11.8 \cr
$M_r<-19.0$ & 11.60$_{-0.06}^{+0.06}$ &  0.26$_{-0.17}^{+0.17}$ 
            & 11.49$_{-0.96}^{+0.78}$ & 12.83$_{-0.08}^{+0.09}$ 
            &  1.02$_{-0.05}^{+0.05}$ & 12.88$_{-0.07}^{+0.08}$ 
            &  1.01$_{-0.01}^{+0.01}$ & 12 &  8.6 \cr
$M_r<-19.5$ & 11.75$_{-0.06}^{+0.06}$ &  0.28$_{-0.18}^{+0.16}$ 
            & 11.69$_{-0.75}^{+0.68}$ & 13.01$_{-0.09}^{+0.09}$ 
            &  1.06$_{-0.06}^{+0.06}$ & 13.06$_{-0.06}^{+0.06}$ 
            &  1.03$_{-0.01}^{+0.01}$ & 12 &  3.9 \cr
$M_r<-20.0$ & 12.02$_{-0.06}^{+0.06}$ &  0.26$_{-0.16}^{+0.16}$ 
            & 11.38$_{-0.93}^{+1.01}$ & 13.31$_{-0.09}^{+0.09}$ 
            &  1.06$_{-0.08}^{+0.08}$ & 13.34$_{-0.07}^{+0.07}$
            &  1.05$_{-0.02}^{+0.02}$ & 12 &  6.2 \cr
$M_r<-20.5$ & 12.30$_{-0.05}^{+0.05}$ &  0.21$_{-0.14}^{+0.16}$ 
            & 11.84$_{-0.42}^{+0.43}$ & 13.58$_{-0.06}^{+0.06}$ 
            &  1.12$_{-0.04}^{+0.04}$ & 13.59$_{-0.06}^{+0.06}$
            &  1.15$_{-0.01}^{+0.01}$ & 12 &  2.8 \cr
$M_r<-21.0$ & 12.79$_{-0.10}^{+0.10}$ &  0.39$_{-0.22}^{+0.20}$ 
            & 11.92$_{-1.27}^{+1.15}$ & 13.94$_{-0.08}^{+0.08}$ 
            &  1.15$_{-0.10}^{+0.09}$ & 13.98$_{-0.06}^{+0.06}$ 
            &  1.28$_{-0.02}^{+0.02}$ & 12 &  2.3 \cr
$M_r<-21.5$ & 13.38$_{-0.16}^{+0.15}$ &  0.51$_{-0.27}^{+0.22}$ 
            & 13.94$_{-0.51}^{+0.48}$ & 13.91$_{-0.84}^{+0.56}$ 
            &  1.04$_{-0.59}^{+0.55}$ & 14.47$_{-0.07}^{+0.07}$ 
            &  1.52$_{-0.05}^{+0.05}$ & 11 &  5.1 \cr
$M_r<-22.0$ & 14.22$_{-0.18}^{+0.13}$ &  0.77$_{-0.18}^{+0.12}$ 
            & 14.00$_{-0.22}^{+0.22}$ & 14.69$_{-0.18}^{+0.21}$ 
            &  0.87$_{-0.52}^{+0.35}$ & 14.88$_{-0.12}^{+0.12}$ 
            &  1.91$_{-0.08}^{+0.09}$ &  7 &  1.0 \cr
\enddata
\tablenotetext{a}{See \S~3 for the HOD parameterization. Mass is in unit 
of $\hMsun$. Error bars are for 1-$\sigma$, which are derived from the 
marginalized distribution.}
\tablenotetext{b}{These are derived parameters: $M_1$, the mass scale of 
a halo that can on average host one satellite galaxy above the luminosity 
threshold; $b_g$, the large-scale galaxy bias factor.}
\tablenotetext{c}{This is the total number of data points (values of $w_p$
plus the number density) used in the fitting.}
\label{tab:bestfit}
\end{deluxetable*}

We perform HOD modeling of the projected two-point correlation function 
$w_p(r_p)$ for each DEEP2 galaxy sample. The jackknife covariance matrices 
estimated from the DEEP2 data are noisy due to the size of the volume probed, 
and we therefore use only the diagonal elements in measuring $\chi^2$. 
In calculating $\chi^2$ for each sample, in addition to $w_p(r_p)$ 
we also include the galaxy number density, obtained by integrating 
the observed luminosity function \citep{Willmer06} and assigning a 
10\% fractional uncertainty to the result.  We apply a Markov Chain Monte 
Carlo (MCMC; see e.g., \citealt{Gilks96}) method to explore the parameter 
space.  For the purpose of comparison, we also model the SDSS galaxy samples 
in \citet{Zehavi05} using the same HOD parameterization and MCMC 
method. The full jackknife covariance matrices are used for the SDSS samples. 
We perform a test with the $M_r<-20.5$ SDSS sample by only including the
diagonal elements of the covariance matrix in the fit and find that the 
best-fit $\chi^2$ is higher, which reflects the fact that the likely 
positive covariance among values of the two-point correlation function in 
adjacent bins is neglected. However, the marginalized distributions of 
HOD parameters are found to be pretty similar to those with the full 
covariance matrix. The test indicates that there are not likely to 
be large systematic uncertainties in the inferred HOD parameters
for the DEEP2 samples for which we use only the diagonal elements of the 
covariance matrices in the fits.

Best-fit HOD parameters for DEEP2 and SDSS samples are listed in 
Table~\ref{tab:bestfit}.  We also list two derived parameters in the table:
$M_1$ (not $M_1^\prime$), the mass scale of a halo that can on average host 
one satellite galaxy above the luminosity threshold, and $b_g$, the large
scale bias factor of galaxies. In this section, we briefly describe the 
modeling results and fits to $w_p(r_p)$, focusing on the DEEP2 galaxies. 
Detailed inspections of individual HOD parameters and 
comparisons between occupation properties of DEEP2 and SDSS galaxies are
presented in the next section.

\begin{figure*}[t]
\epsscale{0.8} 
\plotone{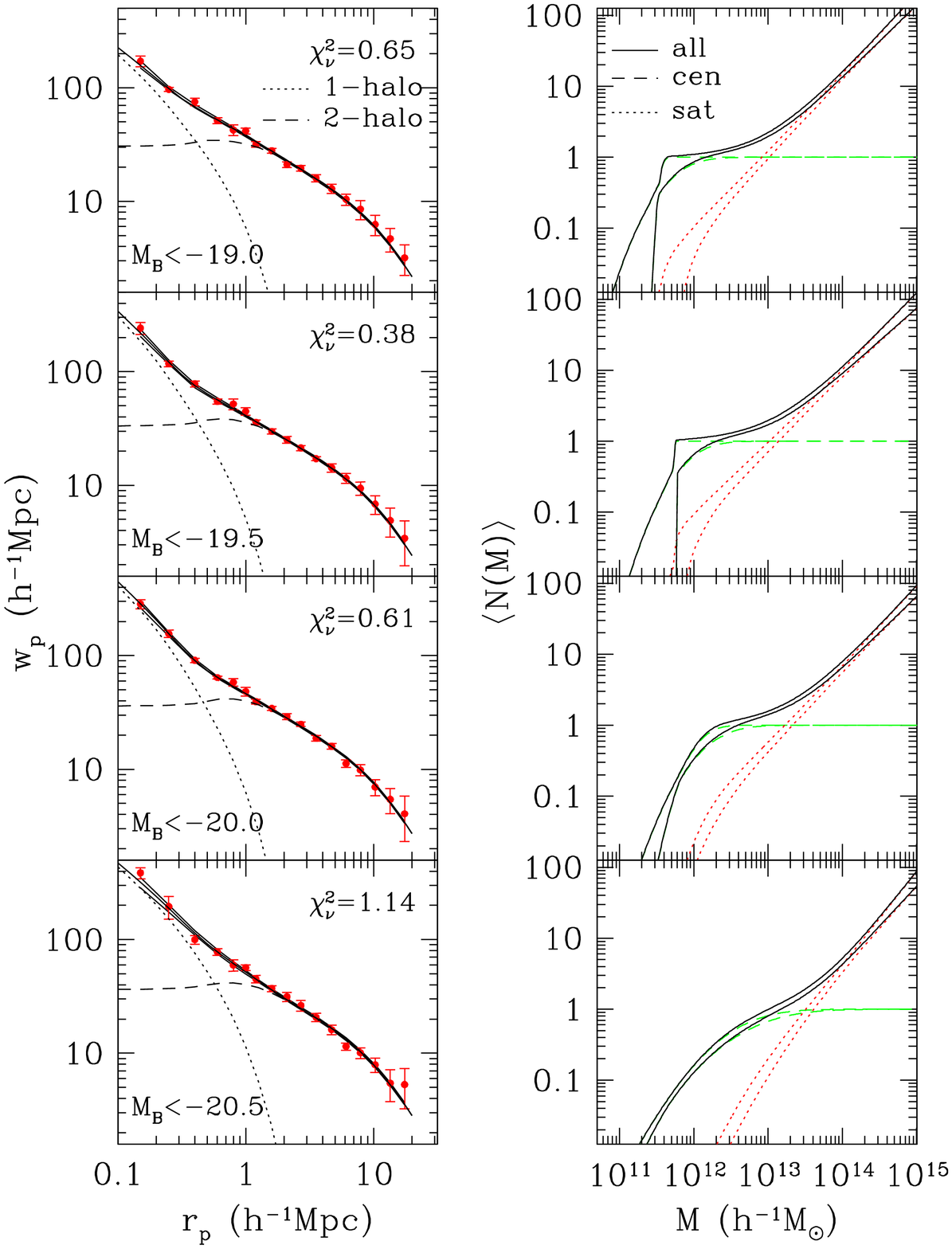}
\epsscale{1.0}
\caption[]{
\label{fig:wp_hod}
HOD fits to $w_p(r_p)$ ({\it left}) and the mean occupation
functions ({\it right}) for DEEP2 galaxy samples. In each 
$w_p(r_p)$ panel, overplotted on the data points are the range of lines 
predicted from models with $\Delta\chi^2<1$ (the reduced $\chi^2$ from the 
best fit is labeled in each panel). The one-halo ({\it dotted line}) and 
two-halo ({\it dashed line}) terms from the best-fit model are also plotted for 
illustration. In each $\langle N(M)\rangle$ panel, the envelopes of mean 
occupation functions from models with $\Delta\chi^2<1$ are plotted
and the total mean occupation function ({\it solid line}) is decomposed into
contributions from central ({\it dashed line}) and satellite ({\it dotted 
line}) galaxies.
}
\end{figure*}

\begin{figure*}[t]
\epsscale{0.8}
\plotone{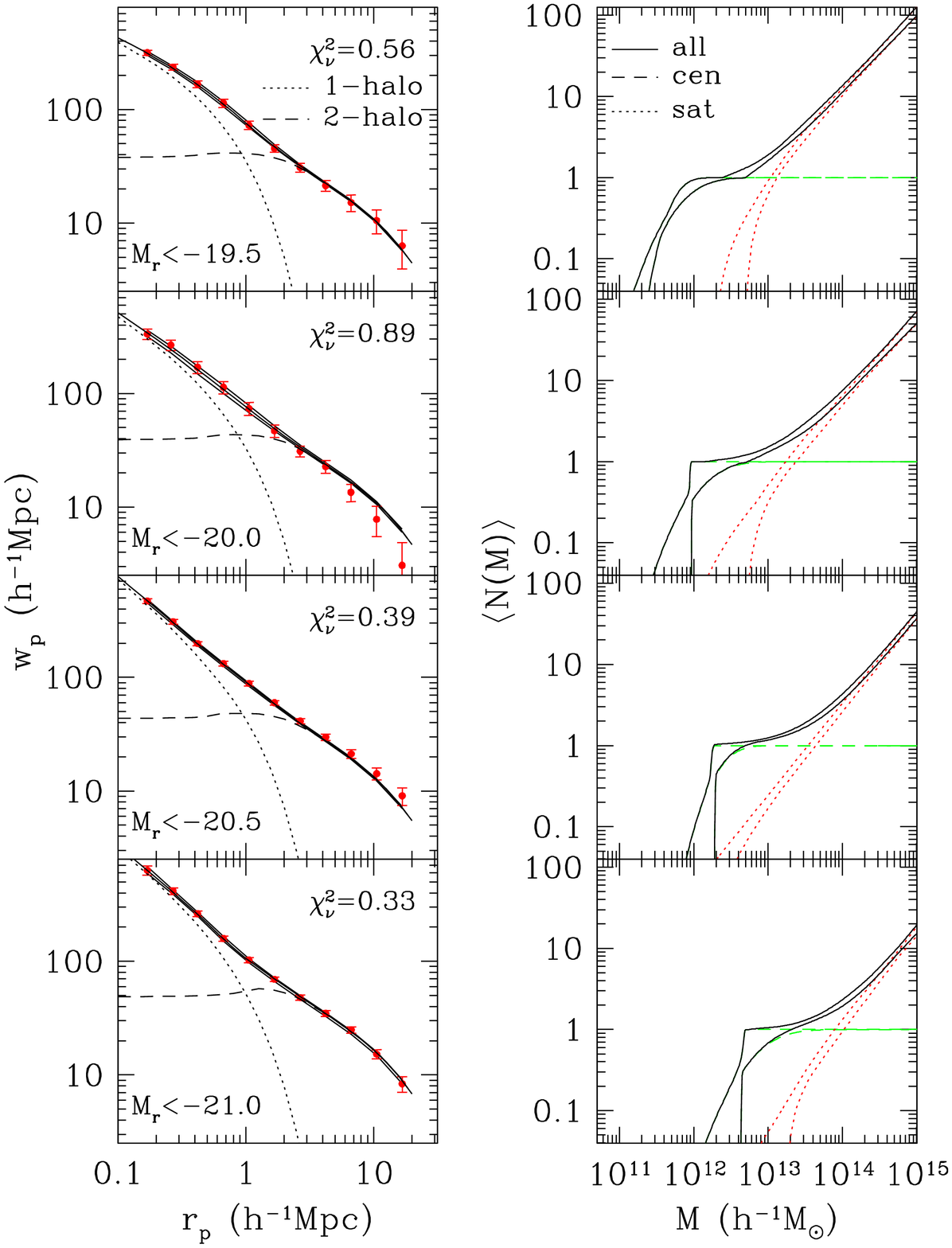}
\epsscale{1.0}
\caption[]{
\label{fig:wp_hod_sdss}
Similar to Fig.~\ref{fig:wp_hod}, but for SDSS galaxy samples.
}
\end{figure*}

\subsection{Results for DEEP2 Galaxies}

The modeling results for the four DEEP2 luminosity threshold samples are 
shown in Figure~\ref{fig:wp_hod}. In each of the left panels, lines from 
model predictions are plotted together with the $w_p(r_p)$ data points.
All fits are excellent. With five parameters (one of which is 
largely determined by the mean galaxy density once the other four are
specified), this success may seem unsurprising. However, it is worth noting
that the combination of the CDM halo distribution (fixed by our assumed
cosmology) and an HOD model cannot fit arbitrary functions --- this is a
physically constrained model with vastly less freedom than, say, a 4th-order
polynomial. Our HOD fits self-consistently produce the two-point correlation
function both on small scales and on large scales.  On small scales, the 
two-point correlation function reflects how galaxies are spatially 
distributed inside halos. On large scales (e.g., $\sim 10\hMpc$), the 
galaxy two-point 
correlation function is simply the matter correlation function multiplied 
by a constant (i.e., the square of the scale-independent galaxy bias factor). 
Our HOD model nicely reproduces the large-scale shape of $w_p(r_p)$, which 
indicates that the matter power spectrum we use has the correct shape on 
these scales.

The central solid line corresponds to the best-fit model and is bounded 
by two solid lines, which are the envelopes of predictions from models
with $\Delta\chi^2<1$.  As we do not have the full error covariance matrix, 
these should only be interpreted as indicative of the associated 1-$\sigma$ 
uncertainty.  The best-fit line is decomposed into contributions
from the one-halo term (intra-halo galaxy pairs; {\it dotted lines}) and 
two-halo term (inter-halo galaxy pairs; {\it dashed lines}), which 
dominate on small and large scales, respectively. The transition from 
the one-halo term to the two-halo term causes an inflection in $w_p(r_p)$, 
leading to departures from a pure power law that are observed for both low 
redshift galaxies \citep[e.g.,][]{Zehavi04,Hawkins03} and high-redshift 
galaxies \citep[e.g.,][]{Ouchi05,Lee05}. If the transition scale is defined 
as the scale where the two contributions to $w_p(r_p)$ are equal, we find that 
it increases from $\sim$0.4$\hMpc$ to $\sim$0.6$\hMpc$ as the luminosity 
threshold of galaxies increases, a manifestation that more luminous galaxies 
reside in more massive halos. 

The mean occupation functions for the four DEEP2 samples are shown in 
the right panels of Figure~\ref{fig:wp_hod}. The mean occupation
function for all galaxies in the sample is decomposed into that for
central galaxies and that for satellites. For each mean occupation
function, we plot the envelope from models with $\Delta\chi^2<1$.
It can be clearly seen that the mean occupation function shifts to 
higher halo mass as the luminosity of galaxies increases. In general,
a low-mass cutoff in the satellite mean occupation function (which is 
usually neglected in the simple three-parameter model) is required by the 
DEEP2 $w_p(r_p)$ data for a good model fit. The best-fit models for the
two brighter samples have a shallow cutoff profile in $\Ncen$. For the 
two fainter samples, the inferred cutoff profile in $\Ncen$ can be either 
sharp or shallow, which leads to the kink seen in the envelope of $\Ncen$
for models with $\Delta\chi^2<1$. The general trend that the central galaxy 
cutoff profile is better constrained for brighter galaxies reflects the fact
that the halo mass function and bias factor change steeply toward the high 
mass end, although the residual redshift-space distortion [because of 
$w_p(r_p)$ being derived from a finite projection] could artificially 
increase the shallowness of the cutoff profile (J. L. Tinker, private 
communication).

We note that the constraints on the HOD models can be further strengthened 
by enforcing the condition that at any halo mass the mean occupation number 
for central or satellite galaxies in a sample with lower luminosity threshold 
is always higher than that for galaxies in a sample with higher luminosity 
threshold. A full implementation of this would require modeling the data 
from all galaxy samples simultaneously, and there would be strong correlations 
among mean occupation functions of different samples.  Here we choose to 
model each sample individually. To roughly account for the above condition,
we drop some models that are in apparent conflict. Specifically,
the mean occupation functions of satellites for the $M_B$$<$$-19.0$ sample 
from a small fraction of models have a cutoff at much higher mass than those 
for the $M_B$$<$$-19.5$ sample, which is not realistic. We therefore only keep 
models for the $M_B$$<$$-19.0$ sample with a satellite cutoff mass 
of $M_0$$<$$10^{12}\hMsun$ for the following discussions. We note that the 
above consistency condition is automatically satisfied by adopting the CLF
method, with the HOD of each sample obtained by integrating the CLF 
over halo mass, although to reach a similar level of flexibility in the HOD
we use here, more parameters need to be introduced in the CLF.

The high-mass end slopes $\alpha$ for mean occupation functions of satellites
are found to be $0.89_{-0.34}^{+0.23}$, $0.97_{-0.11}^{+0.11}$, 
$1.08_{-0.06}^{+0.05}$, and $1.27_{-0.12}^{+0.12}$ in order of increasing 
luminosity threshold, where each value is quoted as the mean with error 
bars from the central 68.3\% of the marginalized distribution (i.e., 
1--$\sigma$ range).  Although the value for the most luminous sample is 
slightly larger, all values are close to unity, which agrees with 
predictions of galaxy formation models \citep[e.g.,][]{Kravtsov04,Zheng05}.

We note that the best-fit HOD model for the brightest DEEP2 sample 
($M_B<-20.5$) has a larger $\chi^2$ than those of the other samples.
Much of the difference between the model and the data happens on small 
scales ($r_p\lesssim 0.5\hMpc$). On these scales, the raw measurements 
of $w_p(r_p)$ are affected by the undersampling of galaxies caused by 
the slitmask target selection algorithm, and they are corrected by using 
the mock galaxy catalogs of \citet{Yan04} \citep{Coil06}. In general, 
the correction increases the amplitude and the slope of $w_p(r_p)$ on 
small scales by a small amount.  As a test of the robustness of our HOD 
results to this correction, we perform an MCMC run with the uncorrected
$w_p(r_p)$ data for this sample and find $\chi^2/{\rm d.o.f.}=0.75$. 
The inferred HOD therefore seems to be robust against such a correction 
--- there is almost no change, except for a slight decrease in the 
number of satellite galaxies in low-mass halos to account for the slightly 
reduced amplitude and slope of the uncorrected $w_p(r_p)$. The slitmask 
correction could in principle be improved by using catalogs generated 
according to the best-fit HOD model and iterating the correction and model 
fitting a couple of times.  However, given the robustness of the best-fit 
HOD under the adopted parameterization, we do not intend to perform such 
iterations in this paper.

We also tried to fit the data with a more flexible HOD parameterization,
which uses a spline line for the satellite mean occupation function 
(similar to what is used in Fig.19c of \citealt{Zehavi05}). We found that
the HOD is adjusted to fit almost every feature in $w_p(r_p)$, and the 
modeling may result in solutions that are not seen in galaxy formation 
models. For instance, for the brightest $M_B<-20.5$ sample, the best-fit 
satellite mean occupation function has an inflection and flattens out 
toward the low halo mass end, which allows it to fit the high amplitude and 
steep slope of $w_p(r_p)$ on small scales that are vulnerable to slitmask
corrections. Therefore, we conclude that this more flexible HOD 
parameterization might be ``over fitting'' the data, and that to robustly 
interpret the present data does not require an HOD form more flexible than 
our five-parameter model.

\subsection{Results for SDSS Galaxies}

The modeling results for SDSS galaxies using the five-parameter model described 
in \S3.2 are shown in Figure~\ref{fig:wp_hod_sdss}. To be concise, only the 
results of the samples that are most relevant to the evolution connection 
between DEEP2 and SDSS galaxies (see \S~\ref{sec:evolutionlink}) are plotted 
in this figure, although we perform HOD modeling for all SDSS samples with 
luminosity cuts ranging from $M_r=-18.0$ to $M_r=-22.0$ (see Table 1).

Unlike the DEEP2 samples, the cutoff profiles of $\NcenM$ for the SDSS 
samples are loosely constrained.  This can be partly explained by noticing that
in the mass range of $10^{11}$--$10^{12}\hMsun$, both the halo mass function
and the halo bias factor at $z$$\sim$1 are steeper than those at $z$$\sim$0,  
and therefore the galaxy number density and the amplitude of the two-point 
correlation function at large scales lead to a better constraint in the 
cutoff profile for DEEP2 galaxies. We also note that complete freedom in 
$\sigM$ [eq.~(2)] leads to a strong correlation between $\Mmin$ and
$\sigM$ in the sense that larger $\Mmin$ corresponds to larger $\sigM$,
which results in poor constraints on $\Mmin$.  We assign a prior $\sigM<0.5$, 
being conservative according to theoretical predictions \citep{Zheng05}, for 
SDSS galaxy samples with luminosity threshold fainter than $L_*$ 
($M_r^*$$=$$-20.44$). 

The transition scales from one-halo term to two-halo term for SDSS samples
are around $1\hMpc$, larger than those for DEEP2 galaxies. The cause of 
such a shift is that the one-halo term for SDSS galaxies appears to be 
shallower than that for DEEP2 galaxies, which again reflects the
change in the slope of the halo mass function in the mass range probed by 
DEEP2 and SDSS galaxies (note that, for the halo definition we adopt, at a 
given mass the comoving sizes of halos at the two epochs are the same). 
This also explains why the departure of the two-point correlation function 
from a power law becomes more prominent at higher redshift (e.g., 
\citealt{Zehavi04,Zheng04,Kravtsov04,Ouchi05}).

\section{Comparisons of the HOD Modeling Results for DEEP2 and SDSS Galaxies}

In this section, on the basis of the MCMC results presented in the
previous section, we compare the HODs of DEEP2 and SDSS galaxies in detail.

\subsection{The Central Galaxy Luminosity as a Function of Halo Mass}
\label{sec:cen}

\begin{figure*}[t]
\plotone{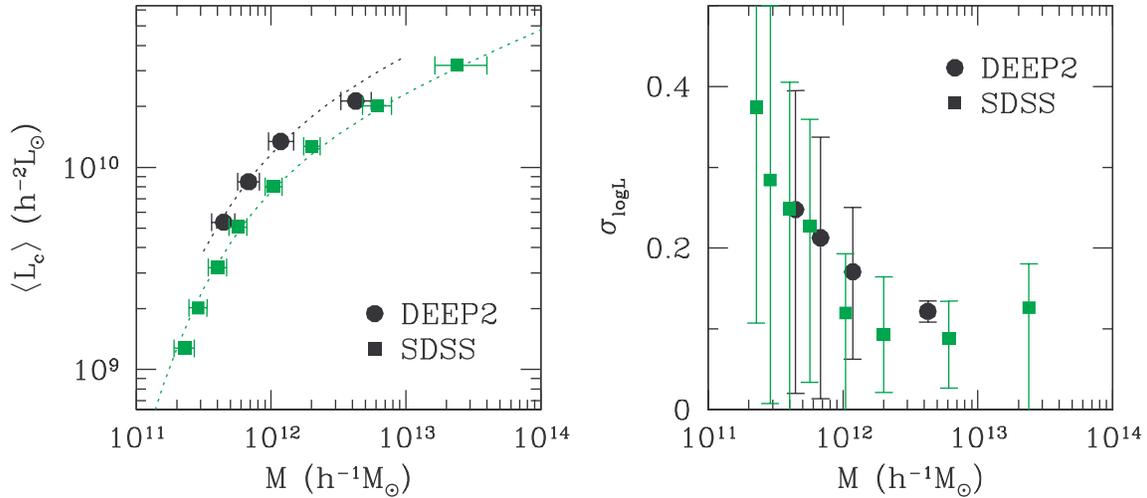}
\epsscale{1.0}
\caption[]{
\label{fig:Lcen}
Distribution of central galaxy luminosity as a function of halo mass.
{\it Left:} Mean luminosity of central galaxies as a function of
halo mass. Filled squares and circles are results for SDSS and DEEP2
galaxies, respectively. The error bars in halo mass indicate the
1--$\sigma$ range of the distribution from MCMC runs. Dotted lines are
calculated using the fitting formula proposed by \citet{Vale06} with
$L_0$ set to $2.8\times10^9\hLsun$ (SDSS) and $4.3\times10^9\hLsun$ (DEEP2)
(see text for details). {\it Right:} Width of the distribution of
central galaxy luminosity as a function of halo mass (see text for details). 
}
\end{figure*}

As mentioned in \S~\ref{sec:hod}, the low-mass cutoff profile of the mean 
occupation function of central galaxies encodes information about the 
distribution of central galaxy luminosities in halos with mass close to
the cutoff mass. By modeling the galaxy two-point correlation functions
for samples with different luminosity thresholds, we can derive the
distribution of the central galaxy luminosity as a function of halo mass.
If the central galaxy luminosity $L_c$ in halos of fixed mass follows 
a log-normal distribution, as suggested by SA and SPH galaxy 
formation models \citep{Zheng05}, then the mass scale $\Mmin$ in 
equation~(\ref{eqn:Ncen}) is the mass
of halos for which the mean central galaxy luminosity $\meanLc$ is 
simply the luminosity threshold of the galaxy sample (as shown in 
\S~\ref{sec:hod}). The left panel of Figure~\ref{fig:Lcen} shows the 
inferred $\meanLc$--$M$ relation for DEEP2 galaxies and for SDSS galaxies.
In the plot, absolute magnitudes are converted to
luminosity in units of solar luminosity by using the Sun's absolute 
magnitudes 4.76 in $r$--band \citep{Blanton03b} and 5.38 in 
$B$--band\footnote{http://www.ucolick.org/$^\sim$cnaw/sun.html} 
(Johnson AB magnitude). 

The $\meanLc$--$M$ relations look similar for the DEEP2 and SDSS galaxies 
except for an offset. At fixed halo mass, there is an offset of $\sim$1.4 
in $\meanLc$ between the DEEP2 and SDSS relations. However, given that the
two samples are selected in different rest-frame bands, this 
offset does not necessarily imply that $z\sim 1$ DEEP2 galaxies are more
luminous than $z\sim 0$ SDSS galaxies at a fixed halo mass. Alternatively, 
at fixed $\meanLc$ there is an offset of $\sim 1.6$ in halo 
mass between the DEEP2 and SDSS samples. Similarly, the interpretation 
of such an offset is not immediately clear. In \S~\ref{sec:evolutionlink} we 
attempt to establish a more meaningful evolutionary link between galaxies 
at the two epochs.

Because of a large range in galaxy luminosity, the SDSS samples probe the 
$\meanLc$--$M$ relation over 2 orders of magnitude in halo mass, from 
$\sim$$2\times 10^{11}\hMsun$ to $\sim$$3\times 10^{13}\hMsun$, as shown in 
the left panel of Figure~\ref{fig:Lcen}. We see that halos with higher mass 
host more luminous central galaxies at both $z\sim 1$ and $z\sim 0$. 
The plot shows that halos of Milky Way size ($\sim$$2\times 10^{12}\hMsun$) 
currently host central galaxies with mean luminosity about 
$L_*$ ($M_r^*=-20.44$, $L_r^*=1.20\times 10^{10}\hLsun$; 
\citealt{Blanton03b}). Higher mass halos begin 
to host galaxy groups and clusters, so we expect that the $\meanLc$--$M$ 
relation changes slope with halo mass. Indeed, the mean luminosity of 
central galaxies increases steeply with halo mass at the low-mass end and 
increases more slowly at the high-mass end, with the slope changing 
continuously from $\sim$2.5 to $\sim$0.3 over the mass range.

The $\meanLc$--$M$ relation from the SDSS modeling results, whose slope
becomes shallower toward higher halo mass, is in general 
agreement with results for local galaxies derived by other methods. 
\citet{Yang05} infer the $\meanLc$--$M$ relation in $b_J$ band through 
associating halos with galaxy groups identified in the 2dFGRS, and they find 
$\meanLc \propto M^{2/3}$ for halos of $M\lesssim 10^{13}\hMpc$ and 
$\meanLc \propto M^{1/4}$ for more massive halos. Using the X-ray masses 
of clusters/groups and the $K$-band luminosity of the brightest cluster 
galaxies in the Two Micron All Sky Survey (2MASS), \citet{Lin04} find
$\meanLc\propto M^{0.26}$ for $M\gtrsim 2\times 10^{13}\Msun$. By modeling
the SDSS galaxy lensing data in \citet{McKay01}, \citet{Cooray05} obtain
the $\meanLc$--$M$ relation for $M\lesssim 10^{13}\Msun$, which can be
approximated as $\meanLc \propto M^{0.75}$. \citet{Vale04} present an
empirical model for the relation between galaxy luminosity and halo mass
by matching the galaxy luminosity function and halo/subhalo mass function 
(see also \citealt{Vale06}) and find that the mass luminosity relation can 
be well approximated by a double power law. \citet{Vale06} advocate
a fitting formula of the following form:
\begin{equation}
\label{eqn:mass_luminosity}
\meanLc=L_0\frac{(M/M_c)^a}{[1+(M/M_c)^{bk}]^{1/k}},
\end{equation}
with $M_c=3.7\times 10^9\hMsun$, $a=29.78$, $b=29.5$, and $k=0.0255$. 
With these parameter values, $L_0$ is the mean galaxy luminosity in halos 
of mass $3.46\times10^{11}\hMsun$.  The relation is steep at low mass and 
becomes $\meanLc\propto M^{0.28}$ at high mass. We find that this fitting 
formula provides a good description of our modeling results of the SDSS 
galaxies if $L_0=2.8\times 10^9\hLsun$ (see the dotted line for SDSS 
galaxies in the left panel of Fig.~\ref{fig:Lcen}).

The DEEP2 galaxy samples we model probe the $\meanLc$--$M$ relation at 
$z\sim 1$ over a smaller mass range, from $\sim$$4\times 10^{11}\hMsun$ to 
$\sim$$4\times 10^{12}\hMsun$. Central galaxies of luminosity $L_*$ 
($M_B^*=-20.7$, $L_B^*=2.56\times 10^{10}\hLsun$, \citealt{Willmer06}), 
which is slightly above the highest luminosity threshold we probe, 
tend to reside in halos a few times more massive than $L_*$ halos
at $z\sim 0$. At $M<2\times 10^{12}\hMsun$, the shape of the $\meanLc$--$M$ 
relation at $z\sim 1$ is consistent with that at $z\sim 0$. At higher 
halo masses, $\meanLc$ appears to increase with $M$ slightly more slowly at 
$z\sim 1$ than at $z\sim 0$.  The dotted line passing through the DEEP2 
measurements in the left panel of Figure~\ref{fig:Lcen} is from the fitting 
formula in equation~(\ref{eqn:mass_luminosity}) with $L_0$ set to be 
$4.3\times 10^9\hLsun$. 

In addition to the mean luminosity of central galaxies, with our modeling 
results, we can also study the width $\sigL$ of the distribution of central 
galaxy luminosity at a fixed halo mass [see eq.~(\ref{eqn:PLc})]. The 
information is encoded in the width $\sigM$ of the cutoff profile of $\NcenM$
and the local slope $p$ of the $\meanLc$--$M$ line ---  
$\sigL=p \sigM/\sqrt{2}$ (see \S~\ref{sec:hod}). The results of the width of 
the central galaxy luminosity distribution are shown in the right panel of
Figure~\ref{fig:Lcen}, where a spline line passing through the $\meanLc$--$M$ 
points is used to infer the local slope $p$ for the SDSS and DEEP2 samples,
independently. The error bar on each point indicates the 1-$\sigma$ scatter in 
$\sigL$ (from its marginalized distribution) at the halo mass. In general,
the scatter is not well-constrained by two-point correlation functions 
for low-luminosity samples, as the halo mass function and halo bias factor 
are not steep at the low-mass end. Therefore, we can remark only on general 
trends in the scatter. We find that at a fixed halo mass, the scatter $\sigL$ 
in the central galaxy luminosity for $z\sim 1$ DEEP2 galaxies is about 
10\%--20\% higher than that for $z\sim 0$ SDSS galaxies, except for the 
brightest DEEP2 sample. For samples at both redshifts, there is a trend that 
$\sigL$ decreases with increasing halo mass, although the significance from the 
modeling results is marginal. In the mass range we can probe, $\sigL$ 
changes from $\sim 0.3$ to  $\sim 0.1$ for both SDSS and DEEP2 galaxies.
The value of $\sigL$ approaches a constant of $\sim 0.11$--0.12 at masses 
above $2\times 10^{12}\hMsun$ for both samples. 

The decrease of $\sigL$ with increasing halo mass (except for the last
SDSS data point), if true, is consistent with theoretical expectations 
\citep[e.g.,][]{Zheng05,Croton06} in the halo mass range probed here. 
The last SDSS data point seems to show an increase but with large uncertainty. 
Although theoretical models also predict that $\sigL$ should increase with halo 
mass again at higher mass, it is at a mass much higher than probed here 
(e.g., $M>5\times10^{13}\hMsun$). 
We note as well that the observed intrinsic scatter of the Tully-Fisher 
\citep{Tully77} relation \citep[e.g.,][]{Giovanelli97}, $\sim 0.20$--0.35 
mag, implies $\sigL=$0.08--0.14 if one assumes that the circular velocity 
is a one-to-one indicator of halo mass. There is also a trend in the
observed intrinsic scatter decreasing as the circular velocity increases,
similar to what we see in our results.  We speculate 
that the reason for the relatively large scatter in central galaxy luminosity 
for lower mass halos may be that the distribution of major star formation 
epochs is broad for these halos.

\subsection{Relation Between Halo Mass Scales of Central and 
Satellite Galaxies}
\label{sec:censat}

\begin{figure*}[t] 
\plotone{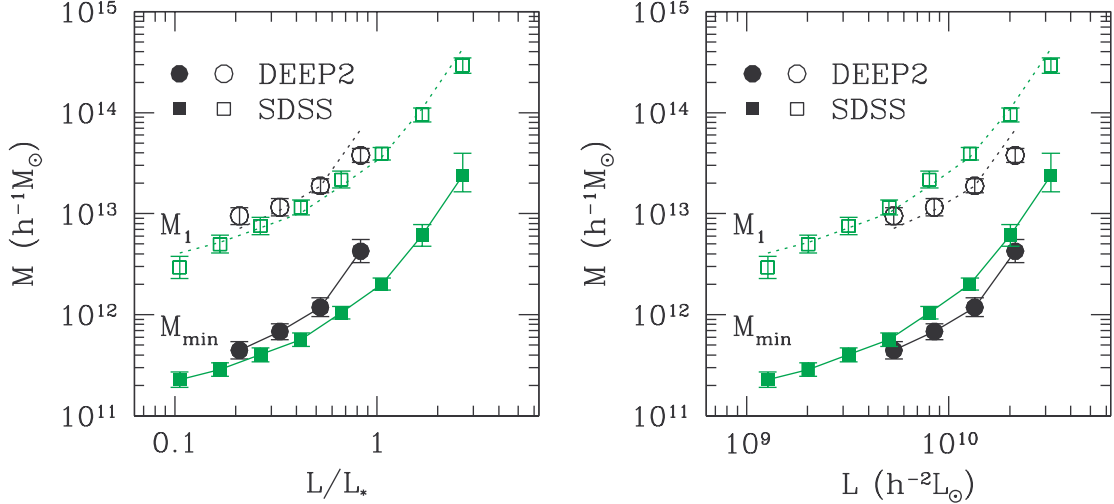}
\epsscale{1.0}
\caption[]{
\label{fig:MminM1}
Mass scales of halos hosting central galaxies and satellites as a function
of galaxy luminosity. 
{\it Left:} Mass scales as a function of $L/L_*$.
Filled symbols connected by solid lines are the relation between the 
characteristic minimum halo mass $\Mmin$ and the luminosity threshold, and
open symbols are that between $M_1$ and the luminosity threshold, 
where $M_1$ is the mass of a halo that on average hosts one satellite galaxy 
above the luminosity threshold. Dotted lines are obtained by multiplying 
the solid lines by a factor of 18 (SDSS) and 16 (DEEP2), respectively.
{\it Right:} Same as the left panel, but the luminosity is expressed
in units of solar luminosity.
}
\end{figure*}

In \S~\ref{sec:cen} we determined the mass scales of central 
galaxies as a function of galaxy luminosity. We now focus on the 
halo occupation of satellite galaxies and the relation between
the halo mass scales of central and satellite galaxies.

In our HOD parameterization, the parameter $\Mmin$ can be interpreted as 
the characteristic minimum mass of the halo hosting a central galaxy above 
the luminosity threshold: $\Ncen=0.5$ at $\Mmin$ [see eq.(~\ref{eqn:Ncen})]. 
For satellite galaxies, the characteristic halo mass scale $M_1$ is
defined as the mass of a halo that on average can host one satellite galaxy 
above the luminosity threshold: $\langle N_{\rm sat}(M_1) \rangle=1$
[note that $M_1$ is different from $M_1^\prime$ in eq.(\ref{eqn:Nsat})]. 
In Figure~\ref{fig:MminM1} we show $\Mmin$ and $M_1$ as a function of
galaxy luminosity for DEEP2 and SDSS galaxies.  In the left panel luminosity 
is relative to $L_*$ at each redshift, and in the right panel luminosity is 
in solar units. For central galaxies, the data points 
are essentially those in Figure~\ref{fig:Lcen} viewed at a different angle.
As mentioned in \S~\ref{sec:cen}, the DEEP2 and SDSS galaxies are selected in 
different rest-frame bands, such that inferring evolution between the samples 
is complex.  Comparing galaxies relative to $L_*$ at each redshift can
roughly account for the different selections and the general dimming of 
galaxies since $z\sim 1$. The plot shows that the host halos of $z\sim 1$ 
$L_*$ galaxies are more massive than those of $z\sim 0$ $L_*$ galaxies,
a trend of downsizing in terms of the host halo mass of ``typical'' galaxies 
with time since $z\sim 1$.

In \citet{Zehavi05}, the HOD modeling for SDSS galaxies reveals a
relation $M_1\sim23\Mmin$. Because of the slightly different HOD 
parameterization, $\Mmin$ and $M_1$ in our analysis do not correspond 
exactly to those in \citet{Zehavi05}. Our modeling results of the SDSS
galaxies, as shown in Figure~\ref{fig:MminM1}, can be expressed as
$M_1\sim 18\Mmin$. \citet{Zheng05} present $M_1$--$\Mmin$ relations in
SA and SPH galaxy formation models for samples of galaxies above varying 
baryonic mass thresholds and find that the scaling factors are about 18 and 
14 for the two models, respectively. Their definitions of $\Mmin$ and $M_1$ 
are more consistent with the ones used here, and this may be the reason 
that our results here are in better agreement with the theoretical 
predictions of \citet{Zheng05}.

The $M_1$--$\Mmin$ scaling relation for $z\sim 1$ DEEP2 galaxies is 
similar to that for $z\sim 0$ SDSS galaxies, as shown in 
Figure~\ref{fig:MminM1}. The scaling factor is about 16, close to the value 
for SDSS galaxies. Dissipationless simulations 
in \citet{Kravtsov04} show a weak trend of the scaling factor decreasing 
towards high redshift, which can be understood as the satellite galaxies 
not having enough time to merge with the central galaxies at higher redshift.

For both data sets, the scaling factor of the $M_1$--$\Mmin$ 
relation has a trend of becoming smaller at the high-luminosity end. High 
luminosity satellite galaxies reside in massive halos. At any given redshift, 
massive halos form late and the accreted satellite galaxies do not have enough 
time to merge with the central galaxies. This effectively reduces the mass
of the halos that on average host one satellite galaxy. Therefore, the effect 
of merging timescale may be the primary reason that the scaling factor 
decreases for very luminous galaxies.

From Figure~\ref{fig:MminM1}, we can also see that at a fixed halo mass 
central galaxies are much more luminous than typical satellite galaxies 
for both DEEP2 and SDSS samples, with the difference being larger toward 
lower halo mass.  This is a manifestation of a CLF bump caused by the 
central galaxy seen in galaxy formation models (\citealt{Zheng05}).

\subsection{Satellite Fraction}

\begin{figure}
\plotone{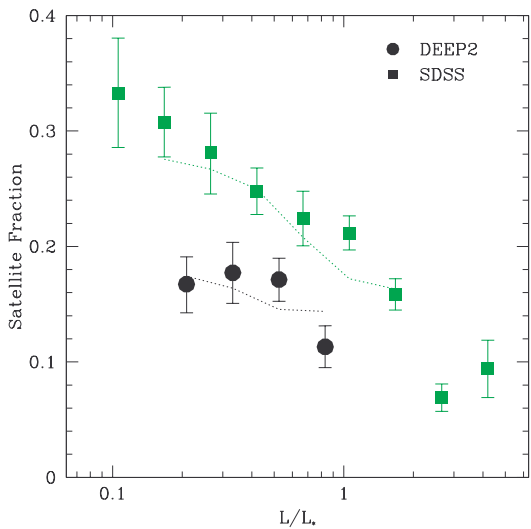}
\epsscale{1.0}
\caption[]{
\label{fig:SatFrac}
Satellite fraction as a function of luminosity threshold for DEEP2 and SDSS
galaxies. The luminosity threshold is expressed in units of $L_*$ at either 
redshift. Symbols show the mean value of the satellite fraction, and the 
error bars indicate the 1--$\sigma$ range of the marginalized distribution 
from MCMC runs. The dotted lines are calculated from interpolating the HOD 
parameters in \citet{Conroy06}, which are inferred through relating halos 
and subhalos in $N$-body simulations to galaxies.
}
\end{figure}

A galaxy at a given luminosity can be either a central galaxy in a 
relatively lower mass halo or a satellite galaxy in a higher mass halo,
with the former being much more probable \citep{Zheng05}. Although at a fixed 
luminosity satellite galaxies are not dominant in number, they play a 
significant role in shaping the two-point correlation function of galaxies
on small scales, where the one-halo term dominates. On these scales, galaxy
pairs are composed of central-satellite and satellite-satellite pairs. 
Therefore, the two-point correlation function can provide strong 
constraints on the overall fraction of satellite galaxies.

Figure~\ref{fig:SatFrac} shows the satellite fraction as a function of
the luminosity threshold for DEEP2 and SDSS galaxy samples based on the MCMC 
runs, where the luminosity threshold is expressed in units of $L_*$. 
SDSS galaxies have a larger satellite fraction than DEEP2 galaxies. 
For example, $\sim$20\% of $L\gtrsim L_*$ SDSS galaxies are
satellite galaxies, compared to $\sim$10\% of $L\gtrsim L_*$ DEEP2 galaxies.
For both DEEP2 and SDSS galaxies, the satellite fraction tends to 
decrease as the luminosity threshold increases.  For $L\gtrsim L_*$, the 
satellite fraction as a function of luminosity seems to drop more steeply. 
This trend with galaxy luminosity reveals that luminous galaxies are much 
more likely to be central galaxies in lower mass halos than satellite 
galaxies in higher mass halos.

\citet{Conroy06} predict the evolution of the luminosity dependence of 
galaxy clustering for the redshift range $0<z<5$ by monotonically relating 
galaxy luminosities to the maximum circular velocity of halos and subhalos 
in $N$-body simulations. Their Table 2 lists the relevant HOD parameters 
at each redshift for various luminosity threshold samples, with different 
number densities. Our inferred HODs at $z\sim 1$ and $z\sim 0$ are in good 
agreement with their non-parametric results. The dotted lines in 
Figure~\ref{fig:SatFrac} are satellite fractions calculated from 
interpolating the parameters in \citet{Conroy06}, which are consistent with 
our results in the relevant luminosity range. We note that satellite 
fractions, obtained here through HOD modeling of the two-point correlation 
function, can in principle be compared to those inferred through 
relating galaxy groups identified in the surveys (e.g., 
\citealt{Gerke05, Coil06b} 
for the DEEP2 survey and \citealt{Berlind06, Weinmann06} for 
the SDSS survey) to dark matter halos for a consistency test.

On large scales, galaxy pairs have contributions from central galaxies 
paired with each other. Being lower in number, satellite galaxies do not 
contribute as much to the two-point galaxy correlation function on large 
scales as on small scales. However, applications that relate
the large-scale galaxy bias factor (relative to dark matter) to the halo bias 
factor often neglect this small contribution from satellite galaxies, which 
can systematically affect the results on the inferred mass scale and number
density of host halos.  In the Appendix we quantify the systematic error
resulting from the ``one galaxy per halo'' assumption using a simple HOD model.

\section{Evolutionary Connections between DEEP2 Galaxies and SDSS Galaxies}
\label{sec:evolutionlink}

Thus far, we have compared different aspects of the halo occupations of 
DEEP2 and SDSS galaxies, treating the results as ``static'' observations of 
the galaxy population at two different redshifts.  We now attempt to 
establish an evolutionary link between galaxies at the two epochs to 
study the evolution of the (statistically) ``same'' galaxies with time.
By ``same'' we mean a galaxy and its most massive progenitor.

HOD modeling converts observed galaxy clustering measurements to a relation 
between galaxy properties and their host dark matter halos. The 
formation and evolution of the halos themselves are dominated by gravity, 
which is well understood from theory and simulations. In principle, we know 
how the halo population at an earlier epoch evolves to that at a later epoch. 
Given the relationships between galaxies and halos at two different epochs, 
the galaxy populations at those epochs can then be linked through the growth 
of their halos.

In what follows, we implement this idea to study the evolution of 
central galaxies in the DEEP2 and SDSS samples.  We first establish a
relationship between $z\sim 0$ halo mass and progenitor halo mass at 
$z\sim 1$ in \S~\ref{sec:halogrowth}. Using this relation and the HOD
modeling results, in \S~\ref{sec:luminosity_evolution} we link the 
$z\sim 0$ central galaxies and their progenitor central galaxies at 
$z\sim 1$ in terms of luminosity to measure the luminosity evolution of 
galaxies since $z\sim 1$.  The stellar mass of a galaxy is perhaps a more 
fundamental physical parameter than luminosity and is more
straightforward to interpret and compare to theoretical models.
Therefore, in \S~\ref{sec:stellarmass_evolution}, using the same halo mass
relationship, we attempt to link central galaxies at the two redshifts in 
terms of stellar mass to study the growth of stellar mass as a function 
of host halo mass.  From this we are able to draw tentative conclusions on 
galaxy evolution and star formation as a function of halo mass during the 
past $\sim$7 billion years. Finally, in \S~\ref{sec:starformation_efficiency}, 
using the stellar mass derived in \S~\ref{sec:stellarmass_evolution}, 
we study how star formation efficiency depends on halo mass at the two 
redshifts.

\subsection{Typical Growth of Halos from $z\sim$1 to $z\sim$0}
\label{sec:halogrowth}

To use our HOD results presented here, what we want to know is the mass 
of the $z\sim 1$ progenitor halo for each of the $z\sim 0$ host halos. 
A dark matter halo merger tree that traces the 
assembly history of halos serves this purpose well.
We use the PINOCCHIO (PINpointing Orbit-Crossing Collapsed HIerarchical
Objects) code developed by \citeauthor{Monaco02a}
\citep[2002a; see also ][]{Monaco02b,Taffoni02} to predict the assembly
history of dark matter halos. PINOCCHIO is based on Lagrangian 
perturbation theory and can generate synthetic catalogs of halos that 
include mass, position, velocity, merger history, and angular momentum.
For the halo assembly history, predictions of PINOCCHIO are much more
accurate than those based on the extended Press-Schechter formalism 
\citep[EPS;][]{Press74,Bond91,Lacey93}. Compared to $N$--body simulations, 
PINOCCHIO can predict the average mass assembly history of halos to 
an accuracy of 10\% or better \citep{LiYun05} with several orders of 
magnitude less computational time. Furthermore, unlike $N$--body simulations,
which need expensive postprocessing to extract merger trees, PINOCCHIO 
directly outputs the merger history of each halo. We perform four realizations 
of PINOCCHIO in a $100{\rm Mpc}^3$ box and three realizations in a 
$50{\rm Mpc}^3$ box to probe assembly histories of halos with mass above and 
below $10^{12}M_\odot$, respectively. For these runs, the grid size is set to
be $128^3$. The mass distribution of the $z\sim 1$ progenitor halos of the
$z\sim 0$ halos is obtained in a series of $z\sim 0$ halo mass bins. The 
code does not include evolution of subhalos within parent halos, but it 
suffices for our purpose here as we focus only on central galaxies.

The results of the typical halo growth from $z\sim 1$ to $z\sim 0$ are 
presented in Figure~\ref{fig:sdss_deep}$c$, which 
shows the relation between $M$, the mass of
$z\sim 0$ halos, and $\Mpro$, the mass of their $z\sim 1$ progenitors. 
The solid line represents the mean mass of the progenitors, and the two 
dotted lines indicate the boundaries of the central 68.3\% distribution 
of the progenitor mass as a function of the $z\sim 0$ halo mass. On average, 
lower mass halos grow earlier, in that more of their final mass is assembled
by $z\sim 1$. The results show that a typical 
$z\sim 0$ halo with $M\sim 3\times 10^{11}\hMsun$ ($M\sim 10^{13}\hMsun$) 
has about 70\% (50\%) of its final mass in place at $z\sim 1$. 
We also use a merging tree code based on the (less accurate) EPS formalism 
and find that it generally predicts late growth for halos and leads to 
$\sim$15\% lower masses of the $z\sim 1$ progenitors in the mass range we
consider. 

\subsection{Luminosity Evolution}
\label{sec:luminosity_evolution}

\begin{figure*}[t]
\plotone{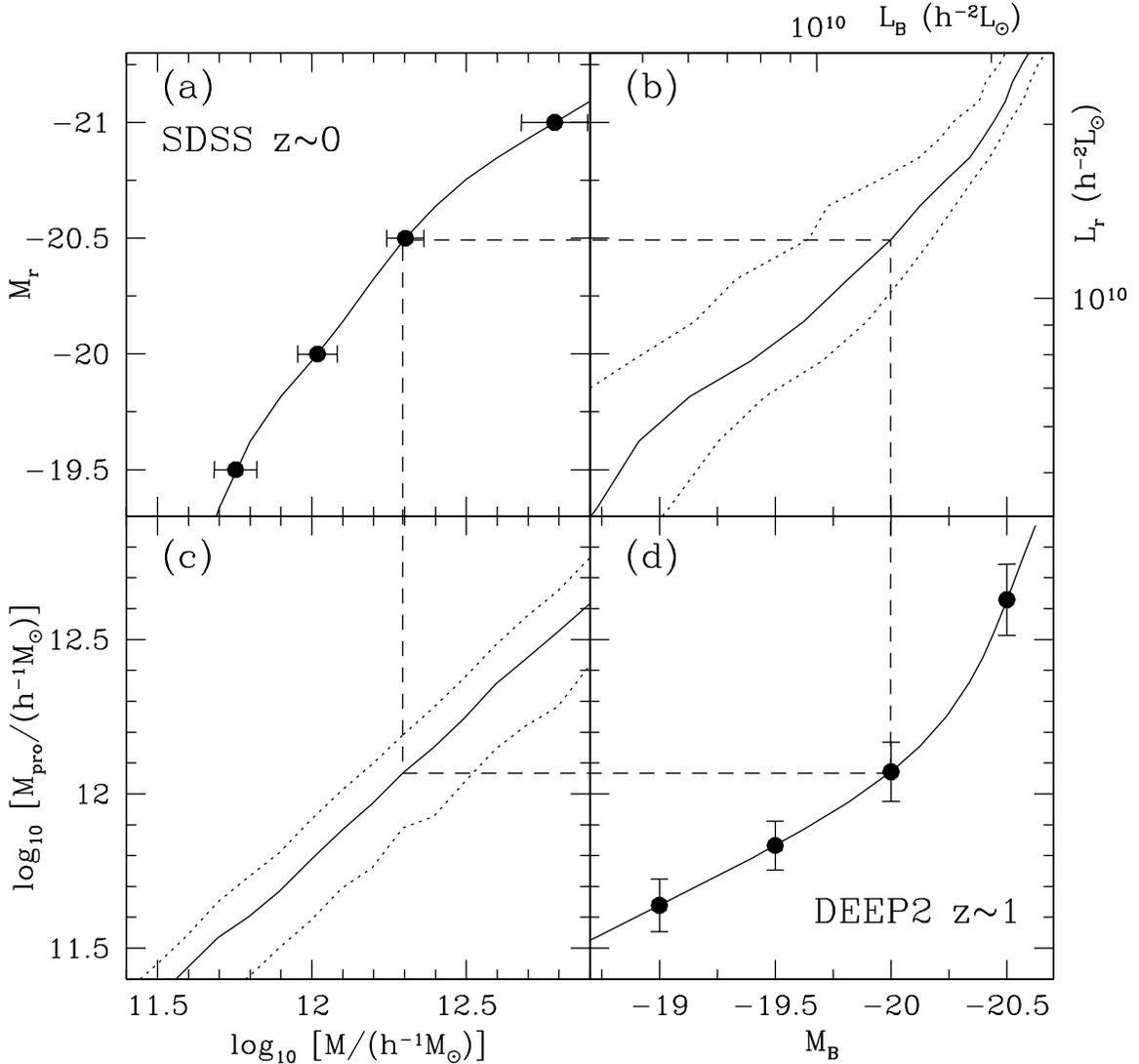}
\epsscale{1.0}
\caption[]{ 
\label{fig:sdss_deep}
Connection between DEEP2 galaxies and SDSS galaxies. 
Panels ({\it a}) and ({\it d})
show the mean luminosity of central galaxies as a function of halo mass at
$z\sim$0 (SDSS) and $z\sim$1 (DEEP2). Panel ({\it c}) shows the relation 
between the mass $M$ of the $z\sim$0 halos and the mass $\Mpro$ of 
their $z\sim$1 progenitors. The solid line is the mean relation, and the
dotted lines mark the central 68.3\% of the distribution. Panels ({\it a}), 
({\it c}), and ({\it d}) lead to a luminosity connection between DEEP2 
galaxies and SDSS galaxies, which is shown in panel ({\it b}). The four 
dashed lines illustrate how this connection is established (see text for 
details).
}
\end{figure*}

The question we attempt to answer is, for galaxies of a given luminosity 
(e.g., $L_*$) at $z\sim 0$, what luminosity did their $z\sim 1$ progenitor 
galaxies have?  We answer this question using the HOD results above and 
the $\Mpro$--$M$ relation shown in Figure~\ref{fig:sdss_deep}$c$.

The $\Mpro$--$M$ relation links halos at $z\sim$1 to those at $z\sim$0.
Together with the $\meanLc$--$M$ relations at these two redshifts derived from 
DEEP2 and SDSS galaxies, which are plotted in Figure~\ref{fig:sdss_deep}$d$ 
and Figure~\ref{fig:sdss_deep}$a$, it enables a connection between DEEP2 
and SDSS central galaxies. The dashed lines across the four panels of 
Figure~\ref{fig:sdss_deep} illustrate how such a connection is established. 
Figure~\ref{fig:sdss_deep}$b$ shows the resultant connection between the 
$r$--band luminosity of $z\sim 0$ central galaxies and the $B$--band 
luminosity of their $z\sim1$ progenitors. For example, the $z\sim1$ 
progenitors of the 
$z\sim0$ $L_*$ ($M_r^*=-20.44$; $L_r=1.20\times 10^{10}\hLsun$) central 
galaxies on average have a $B$--band luminosity of $M_B=-20.0$ 
($L_B=1.34\times 10^{10}\hLsun$), less than  $L_*$ at $z\sim 1$.

The above connection is between luminosities in two different rest-frame 
bands at the two redshifts. A more interesting comparison may be between 
luminosities in the same rest-frame band. The best way to achieve such a goal 
is to have galaxy samples observed in the same rest-frame band, which is not 
the case for samples analyzed here and is hard to achieve in general at 
different redshifts.  However, we can obtain a rough connection by making a 
transformation between luminosities in the two bands. We use the 
publicly-available {\tt kcorrect} code \citep{Blanton03a} to estimate the 
median $M_B$ values (where $B$ is a Bessell AB magnitude) for SDSS galaxies 
in a series of narrow $M_r$ bins, to facilitate comparison with $M_B$ samples 
in the DEEP2 data. Using the SDSS DR4 data and the same magnitude and 
redshift thresholds as \cite{Zehavi05}, we find that the median $M_B$ value 
for galaxies at a fixed $M_r$ is $M_B\simeq M_r+1$. 
Our transformation implies that $z\sim 0$ $L_r^*$ 
galaxies have a median $L_B\sim 8.0\times 10^9\hLsun$. On average their
$z\sim1$ progenitors have $M_B=-20.0$ (Fig.~\ref{fig:sdss_deep}$b$), 
i.e., $L_B=1.34\times 10^{10}\hLsun$. It is not entirely surprising that their
progenitor galaxies are more luminous, as stars fade as they age.  
Using the \citet{Bruzual03} model, 
we find that the amount of passive luminosity evolution at a fixed time 
interval depends on the age of the stellar population, in the sense that 
younger populations fade more. This is related to the fact that luminous 
massive stars have a shorter lifetime.
For a stellar population with 0.2--1 solar metallicity, which 
forms its stars in a single instantaneous burst at redshift $z=$1.5, 2, 2.5, 
or 3, its rest-frame $B$--band luminosity would decrease by about 1.8, 1.4, 
1.2, or 1.0 mag, respectively, from $z=1$ to $z=0$. Using these values 
as an estimate, a $z\sim1$ progenitor central galaxy would have 
$L_B \sim$2.6--5.3$\times10^9\hLsun$ (higher luminosity for stars forming at 
higher redshift) if they were passively evolving to 
$z\sim 0$. This is fainter than the $z\sim 0$ central galaxies 
($L_B\sim 8.0\times 10^9\hLsun$), which implies that additional stars must 
have been added to the system since $z\sim 1$, through star formation and/or 
galaxy mergers.

\subsection{Stellar Mass Evolution}
\label{sec:stellarmass_evolution}

\begin{figure*}
\plotone{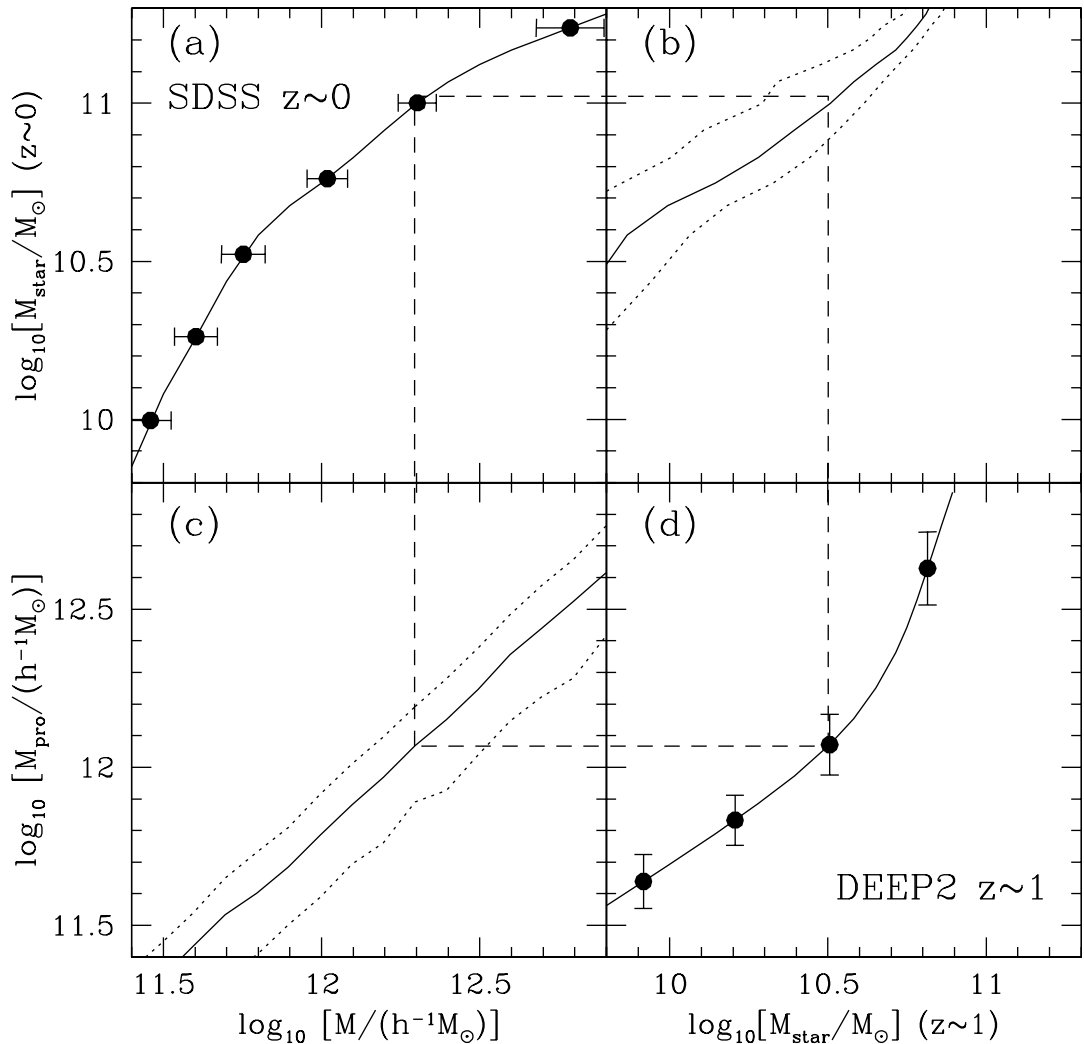}
\epsscale{1.0}
\caption[]{
\label{fig:Mstar}
Similar to Fig.~\ref{fig:sdss_deep}, but in terms of
mean stellar mass. 
}
\end{figure*}

The connection between luminosities of local central galaxies and their 
progenitors provides interesting information on the evolution of the stellar 
component of galaxies from $z\sim 1$ to $z\sim 0$. However, a comparison in 
terms of stellar mass would be more informative and straightforward to 
interpret, as stellar mass is a more fundamental physical parameter than 
luminosity. As with luminosities, the ideal way to make this comparison is to 
construct stellar-mass-selected galaxy samples and to perform HOD modeling 
of the observed clustering of these samples, which we reserve for a future 
study. With the luminosity-selected samples used here, we use a simple 
estimation of stellar mass derived from the galaxy luminosity and color.  
While the correlation between stellar mass and galaxy luminosity can have a 
fair amount of scatter, we attempt here to extract useful information from 
the {\it mean} relation.  The following analysis is presented as a proof of 
concept as to how one could attempt to study evolution of galaxies 
in terms of stellar mass and as a function of halo mass, and it serves only 
as a useful first-order approximation for analyses of stellar-mass-selected 
samples. With these caveats in mind, we proceed as follows.

For the DEEP2 galaxies, stellar masses are derived by \citet{Bundy05} 
for the subset for which $K_s$-band imaging exists, assuming a Chabrier 
stellar initial mass function (IMF; \citep{Chabrier03}).
An empirically-derived relation between rest-frame $UBV$ colors, redshift,
and stellar mass is then used to estimate 
masses for the rest of the DEEP2 sample (C. N. A. Willmer, private 
communication). For the SDSS samples, we estimate the stellar mass from the 
$g-r$ color and the $r$--band luminosity using the relation given by 
\citet{Bell03}. The dominant source of uncertainty in estimating stellar 
mass using the above method is the IMF of stars \citep{Bell01}. The ``diet''
Salpeter IMF \citep{Bell03} is used for the SDSS galaxies. The effect
of different IMFs is largely an offset in the estimated stellar mass.
For example, stellar mass with the diet Salpeter IMF (Chabrier IMF) is 
70\% (50\%) that with the Salpeter IMF. We multiply the DEEP2
stellar masses with the Chabrier IMF by a factor of 1.4 to convert them to a 
diet Salpeter IMF,  so that stellar masses from both surveys can be 
compared. 

\begin{figure}
\plotone{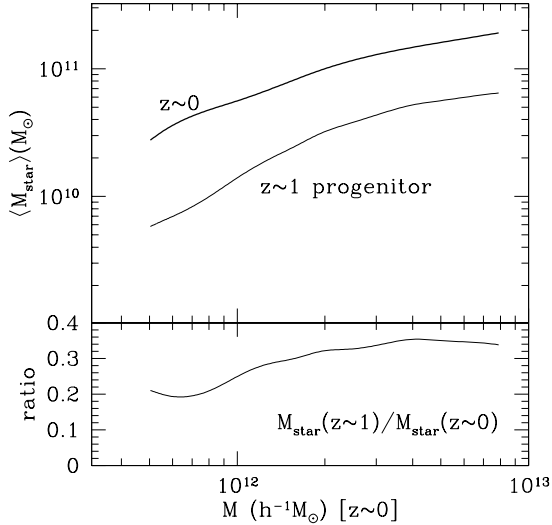} 
\epsscale{1.0}
\caption[]{
\label{fig:MstarRatio}
Mean stellar mass in $z\sim 0$ central
galaxies and that in their $z\sim 1$ progenitors ({\it top}) and their ratio 
({\it bottom}) as a function of the $z\sim 0$ host halo mass.
}
\end{figure}

The mean stellar mass of galaxies and the scatter are calculated in a series 
of narrow luminosity bins. We can then convert the relation between central 
galaxy luminosity and halo mass to that between stellar mass of central 
galaxies and halo mass.  With the present-day central galaxies and their 
progenitors at $z\sim 1$ linked through the $\Mpro$--$M$ relation, we recast 
Figures~\ref{fig:sdss_deep} in terms of stellar mass, which is shown in 
Figure~\ref{fig:Mstar}. Here Figure~\ref{fig:Mstar}$b$ shows the 
relation between the stellar mass of $z\sim 0$ central galaxies and 
that of the progenitors at $z\sim 1$.  In Figure~\ref{fig:MstarRatio}, the 
(arithmetic) mean stellar masses in central galaxies at $z\sim 0$ and those 
of their $z\sim 1$ progenitors as a function of the {\it present-day} halo 
mass are plotted, which allows an estimate of the average growth of
central galaxy stellar mass from $z\sim 1$ to $z\sim 0$ as a function of 
halo mass.  Limited by the luminosity range of the 
data, the halo mass range that can be probed is slightly greater than 1 
order of magnitude. The bottom panel of Figure~\ref{fig:MstarRatio} shows 
the ratio of the mean stellar masses in the $z\sim 1$ progenitors and the 
$z\sim 0$ central galaxies as a function of the present-day halo mass, which 
is expected to be insensitive to the choice of the IMF. 
At $5\times 10^{11}\hMsun$, on average, a central galaxy had $\sim$20\% of 
its present stellar mass in place at $z\sim 1$. The ratio gradually 
increases to $\sim$33\% around $2\times 10^{12}\hMsun$ and is roughly constant 
up to the highest halo mass we can probe. 
For galaxies that reside in $2\times 10^{12}\hMsun$ halos 
(i.e., $z\sim 0$ $L_r^*$ galaxies), the increase in stellar mass from 
$z\sim 1$ appears to be consistent with the luminosity 
comparison presented in \S~\ref{sec:luminosity_evolution} 
if most stars in their $z \sim 1$ progenitors do not form at a very high 
redshift and the luminosity is roughly proportional to stellar mass.
As shown in Figure~\ref{fig:Mstar}$c$, at $z\sim 1$, progenitor 
halos have already reached more than 50\% of their total present-day mass.
Therefore, in the mass range probed here, halos have most of their mass
assembled by $z\sim 1$ but have most of their central galaxy stars 
assembled/formed more recently. The mass scale of 1--2$\times 10^{12}\hMsun$
for present-day halos appears to be a transition scale, below which
a relatively larger fraction of stars have been added to the central 
galaxies during the period from $z\sim 1$ to $z\sim 0$.

In general, there are two processes that can add stellar components to 
central galaxies. One is star formation, and the other is galaxy merging, 
either merging of satellites with central galaxies or merging of central 
galaxies in different halos. The star formation includes that which occurs in 
the central galaxies and the star formation in the satellite galaxies that
eventually merge onto central galaxies. If we know the halo occupation as a 
function of stellar mass at $z\sim 1$, we can evolve the subhalo and halo 
population at $z\sim 1$ to $z\sim 0$ using high-resolution $N$-body 
simulations or an analytic code \citep[e.g.,][]{Zentner05}, assuming no star 
formation.  Such a calculation would give the amount of stars acquired 
through mergers. The difference between the evolved stellar mass occupation 
and the true stellar mass occupation in halos of a given mass (inferred from 
$z\sim0$ data) would tell us the amount of stars formed during the period 
from $z\sim 1$ to $z\sim 0$. While the samples we model are not well suited 
for such a sophisticated analysis, it is a goal we plan to pursue in future 
work. 

The results presented in Figure~\ref{fig:MstarRatio} nevertheless allow us 
to place interesting limits on the amount of stars added to the 
present-day central galaxies by star formation and merging, in an average
sense. As seen from Figures~\ref{fig:sdss_deep} and \ref{fig:MstarRatio}, 
for $z\sim 0$ halos with $M=5\times10^{11}\hMsun$, their $z\sim 1$ 
progenitors have assembled $\sim$70\% of the $z\sim 0$ mass while the 
progenitor central galaxies have $\sim$20\% of the stellar mass in place. For 
low-mass halos, it is reasonable to assume that the stellar mass in the 
progenitor halo is dominated by those 
from central galaxies (see, e.g., \citealt{Zheng05}). 
The progenitor halo therefore assembles the remaining $\sim$30\% mass by 
$z\sim 0$ through smooth mass accretion and/or merging with one or more smaller 
halos. If the fraction of stellar mass inside this 
additional mass is the same as in the progenitor halo (i.e., smaller 
halos are assumed to have the same star-formation efficiency as the 
progenitor of the $5\times10^{11}\hMsun$ halo), then at most 
$\sim$9\% of the final stellar mass can be gained by this process. That is, the 
stellar mass in the progenitor galaxy and that in the smaller halos that 
could merge with it or the mass that is accreted 
can only amount to $\sim$30\% of the final stellar mass in the 
$z\sim 0$ central galaxy.
Therefore, for central galaxies in $z\sim 0$ halos with 
$M=5\times 10^{11}\hMsun$, $\sim$70\% of the stars should form between 
$z\sim 1$ and $z\sim 0$.

Similar reasoning can be applied to higher mass halos, but one should keep 
in mind that the contribution of stellar mass in satellites increases with 
halo mass.  For $z\sim0$ halos with the highest mass we can 
probe ($M\lesssim10^{13}\hMsun$, which host galaxy groups), satellite galaxies
contribute a substantial amount of the stellar mass in the halo. 
The $z\sim 1$ progenitor of an average halo of this mass has assembled 
$\sim$54\% of the total mass, and the progenitor central galaxy has  
$\sim$33\% of its stellar mass assembled.  Neglecting star formation, the
progenitor central galaxy can merge with other central galaxies (with
their total halo mass amounting to the remaining $\sim$46\% of the $z\sim 0$ 
halo mass), and the total contribution to the stellar mass in the final 
$z\sim 0$ central galaxy from all the merged progenitor central galaxies 
would be at most 60\%. Merging of satellites to the final central galaxy 
would make an additional contribution to the stellar mass.  We crudely estimate
this contribution as follows. In a halo, satellite galaxies are usually 
fainter than the central galaxy, and in galaxy groups, the brightest 
satellites are likely to be $\sim$1.6 mag ($\sim$25\%) fainter than 
the central galaxies (i.e., the luminosity gap; see \citealt{Milosavljevic06} 
and references therein). This means that even if all the brightest satellites
in halos that are assembled into a $z\sim0$ $10^{13}\hMsun$ halo are able
to merge onto the final central galaxy, their contribution to the stellar 
mass is only about 25\% of that from merging of central galaxies. Thus, the 
total stellar mass from merging of central galaxies and satellites may 
likely account for $\sim 60\%+60\%\times25\%=75\%$ of that in the final 
central galaxies.  The remaining $\sim$25\%  would then be the result of 
star formation between $z\sim 1$ and $z\sim 0$.

\begin{figure}
\plotone{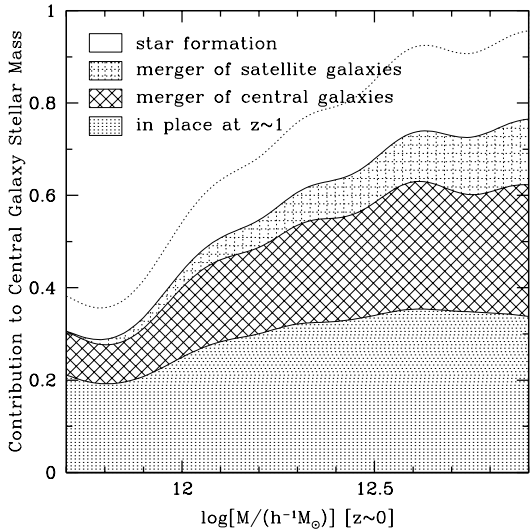}
\epsscale{1.0}
\caption[]{
\label{fig:contribution}
Illustration of contributions to the average growth (since $z\sim 1$) of 
stellar mass of central galaxies residing in $z\sim 0$ halos, as a function 
of the $z \sim 0$ halo mass.  See text for more details of the 
estimation based on the HOD modeling of galaxy clustering for the DEEP2 and 
SDSS galaxy samples. The top solid line would move up to be the dotted 
line after an upward correction of 25\% for the DEEP2 stellar mass 
estimation (see text).
}
\end{figure}

Using this same line of reasoning, we illustrate in 
Figure~\ref{fig:contribution} the different contributions to the $z\sim0$ 
central galaxy stellar mass in the halo mass range we are able to probe. 
In this plot, a linear interpolation in logarithmic halo mass, based on the 
above limits for the low- and high-mass ends, is assumed to estimate the 
contribution from satellite mergers.
Based on these crude estimations, we find an interesting result:  on
average, a large fraction of stars in $z\sim 0$ central galaxies residing
in low-mass halos formed since $z\sim 1$ (e.g., $\sim$70\% for $5\times 
10^{11}\hMsun$ halos), while only a small fraction of stars formed for 
central galaxies in high-mass halos (e.g., $\sim$25\% for 
$\lesssim 10^{13}\hMsun$ halos) over the same period.  
This trend seems to be a manifestation of the so-called downsizing
-- a pattern in which the sites of active star formation shift from high-mass
galaxies at early times to lower-mass systems at later epochs 
\citep[e.g.,][]{Cowie96,Juneau05}.  If the trend continues to 
higher halo mass, beyond the luminosity range we probe in this paper, 
there may be no substantial star formation occurring between $z\sim 1$ 
and $z\sim 0$ in very massive halos. 

Are these estimates reasonable? We can compare them to values obtained
from the global cosmic star formation history (e.g., \citealt{Madau96}),  
probed through different techniques. Using a recent compilation of data in 
\citet{Fardal06}, we estimate that about 40\% of all the stars in the present 
universe formed between $z=1$ and $z=0$. Based on stellar mass estimates
from the COMBO-17 survey, \citet{Borch06} find that the total stellar mass 
density of the universe has roughly doubled since $z\sim 1$. These numbers 
are well bracketed by our values. 

The mass dependence of star formation that we find here can also be compared 
with other results.  We find that the stellar mass growth rate of central 
galaxies depends on the host halo mass.  Since halo mass and central galaxy 
stellar mass are closely related, this implies that the growth rate depends 
on stellar mass itself.  \citet{Jimenez05} use spectroscopic modeling to 
infer the star formation histories of individual SDSS galaxies and 
derive the mean star formation history as a function of galaxy stellar 
mass. By integrating the fraction of stellar mass in the last 6.8 billion 
years as a function of stellar mass (sum of the dotted and all the solid 
lines in the top panel of their Fig.4), we find that 
for $z\sim 0$ galaxies with stellar mass $\sim2\times 10^{10}\Msun$, 
$\sim$70\% of their stars formed between $z\sim0.8$ and $z\sim 0$.  This 
fraction drops to $\sim$40\% for galaxies with a $z\sim 0$ stellar mass of 
$\sim2\times 10^{11}\Msun$. Using the star formation histories of SDSS 
galaxies estimated by \citet{Panter06}, which improves on the modeling 
in \citet{Jimenez05}, the above numbers become $\sim$60\% and $\sim$25\%. 
Our estimates show reasonable qualitative agreement with these results.
It is worth noting that our method differs significantly from that of
\citet{Panter06},  who use stellar 
population modeling of $z\sim 0$ SDSS galaxies with no reference
to $z\sim 1$ galaxies, and do not use the clustering of either population.
Our present analysis uses the population synthesis model only to obtain
stellar mass-to-light ratios (not evolutionary ages), and our conclusions
about stellar mass growth result from the HOD modeling of clustering
at the two different redshifts. The agreement between the results from the
two methods is therefore impressive.

Our results can also be compared with studies of the evolution of star 
formation rates (SFRs) as a function of stellar mass.
Analyzing the AEGIS (All Wavelength Extended Groth Strip 
International Survey) data out to $z\sim 1$, \citet{Noeske07} find 
that star-forming galaxies form a distinct sequence with 
a power-law dependence of the SFR on the stellar mass (see also 
\citealt{Noeske07b}). 
Converted to the ``diet'' Salpeter IMF, their result can be written as
$SFR=3.6\Msun~{\rm yr}^{-1}\times (M_{\rm star}/10^{10}\Msun)^{0.67}$ for 
$M_{\rm star}$ in the range of 1.4--14$\times 10^{10}\Msun$ and $0.2<z<0.7$. 
If we parameterize the total stellar mass growth rate from both star 
formation and mergers as a power-law, 
$\dot{M}_{\rm star}=f(t)M_{\rm star}^\gamma$ (i.e., with a time-dependent
normalization and time-independent power-law index), we can then solve for the
mean relation averaged over $0<z<1$ utilizing the stellar mass
connection between these two epochs (i.e., Fig.~\ref{fig:Mstar}$b$).
For $z\sim 0$ central galaxies with stellar mass in the range of 
0.7--5$\times 10^{10}\Msun$, we
find that the mean growth rate of their stellar mass during the past 
$\sim$7Gyr is $\dot{M}_{\rm star}=2.9\Msun~{\rm yr}^{-1}\times 
(M_{\rm star}/10^{10}\Msun)^{0.67}$. Compared with the SFR in the star 
formation phase mentioned above, it is interesting to notice that the
general growth has the same dependence on the stellar mass but with only 
a $\sim$20\% lower normalization, which suggests that, for these galaxies, 
star formation is the dominant mode of stellar mass growth and the star 
formation phase on average may occupy a substantial fraction of 
the 7 Gyr interval. Modeling galaxy clustering data in more redshift 
slices between $z\sim 1$ and $z\sim 0$ would lead to a better understanding 
of the implication of the general growth rate.

The numbers derived here should be taken as a first-order estimate, as we
have performed an approximate calculation as a proof of concept.
There are two limiting factors in the current data. As mentioned earlier, 
the samples used here are defined using galaxy luminosity, and modeling 
stellar mass-selected samples would be more appropriate for measuring the
evolution of the stellar mass growth rate as a function of halo mass.
The other complication is the different rest-frame color selections of the 
galaxies, in particular the effect that the DEEP2 samples used are not 
entirely volume limited for red galaxies (see \S~\ref{subsec:deep2}).  
This could lead to an underestimate of the mean stellar mass, as red 
galaxies have a higher stellar mass than blue galaxies at a given luminosity. 
Therefore, we may be overestimating the stellar mass evolution from 
$z\sim 1$ to $z\sim 0$. With smaller, volume-limited DEEP2 samples at a 
slightly lower redshift, we estimate that the fraction of galaxies missing 
in each DEEP2 sample is $\sim$10\% and we also find that
the mean stellar mass is likely to be underestimated by 20\%--30\%. 
This means that $\sim$25\% more of the stellar mass could have been 
in place in the $z\sim 1$ progenitors than the numbers we quote here, which
would move the line in the bottom panel of Figure~\ref{fig:MstarRatio} 
(and the lowest line in Fig.~\ref{fig:contribution}) up by $\sim$25\%.
Consequently, the top solid line in Figure~\ref{fig:contribution} would 
shift upward to the place of the dotted line. The star formation contribution 
to the stellar mass growth at the lower (higher) halo mass end we probe 
would change from $\sim$70\% ($\sim$25\%) to $\sim$65\% ($\sim$5\%). 
However, despite these uncertainties, the general trend of the contribution 
to the stellar mass growth as a function of host halo mass remains the same. 
Therefore, the halo mass-dependent evolution trend inferred from the data 
is a robust result.

\subsection{Evolution of the Star Formation Efficiency}
\label{sec:starformation_efficiency}

\begin{figure}
\plotone{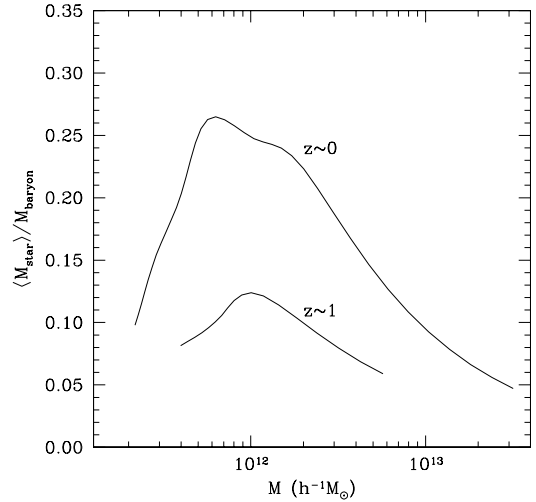}
\epsscale{1.0}
\caption[]{
\label{fig:MstarFrac}
Mean fraction of baryons in host halos that are converted into stars in 
central galaxies for $z\sim 0$ (SDSS) and $z\sim 1$ (DEEP2) galaxies.
}
\end{figure}

With the relation between stellar mass and halo mass at two epochs in 
hand, we now present the evolution of the star formation efficiency 
and discuss the implications.  

Assuming that the baryon fraction in halos equals the mean baryonic fraction 
$f_b$ in the universe, for which we adopt the value 15.7\% (see 
\S~\ref{sec:cosmology}), then stellar mass can be used to study the star 
formation efficiency as a function of halo mass at $z\sim 0$ and $z\sim 1$.  
Specifically, we associate a halo of mass $M$ with a baryon mass $M_b=f_bM$ 
and define $M_{\rm star}/M_b$ as the star-formation efficiency, where 
$M_{\rm star}$ is the stellar mass of the {\it central} galaxy in the halo. 
This star formation efficiency is, in essence, the value integrated from 
the redshift of formation to the epoch in consideration ($z\sim 1$ 
for DEEP2 galaxies and $z\sim 0$ for SDSS galaxies) and reflects the 
fraction of baryons associated with halos that have been converted to 
stars in the central galaxies by the given epoch. In practice, 
the baryon fraction in a halo may not equal the global fraction, due to
various physical processes.  For example, in a low-mass halo there may be a 
smaller fraction of baryons accreted because of the shallow potential 
well.  Therefore, more accurately, the halo star formation efficiency 
defined here is the true star formation efficiency times the baryon accretion
efficiency of the halo. Nevertheless, it is still an instructive quantity 
that can be interpreted as an apparent star formation efficiency.
We note that by definition this efficiency does not include stars
in satellite galaxies, whose contribution increases with halo mass
(about a few percent at $5\times 10^{11}\hMsun$ and $\sim$40\% 
at $10^{13}\hMsun$, according to the estimates in 
\S~\ref{sec:stellarmass_evolution}).

Figure~\ref{fig:MstarFrac} shows this star formation efficiency, reflecting 
the fraction of baryons that are converted into stars at the two epochs.
Both lines are calculated from the ratio of the arithmetic mean stellar mass
of central galaxies in halos of a given mass to the average baryon mass 
associated with these halos.  At $z\sim 0$, there is a characteristic mass 
scale of halos, $\sim 6\times 10^{11}\hMsun$, where the average conversion 
efficiency from baryons to stars in central galaxies reaches a maximal value.
Even at this peak, only $\sim$27\% of the baryons in the halo are converted 
into stars of the central galaxy.  The star formation efficiency drops 
steeply at lower halo masses and declines more slowly at higher halo masses. 

At $z\sim 1$, our inferred average conversion efficiency has a similar 
behavior to that at $z\sim 0$, but the overall conversion efficiency 
decreases and the peak is reached at a larger halo mass 
($\sim$$10^{12}\hMsun$) with a value of $\sim$12\%. While the absolute 
value of the efficiency depends on the IMF, the trend as a function of mass 
and thus the characteristic halo mass should not. 

\citet{Shankar06} infer the star formation efficiency at $z\sim 0$
by matching the observed galaxy stellar mass function with the theoretical 
halo mass function. Their result (see the dashed line in their Fig.5) is in 
a good agreement with ours on the trend of the efficiency with halo mass;
our values are slightly lower than theirs but are well within the error bars. 
\citet{Mandelbaum06} perform HOD modeling of galaxy-galaxy lensing for 
stellar  mass-selected SDSS galaxy samples. The stellar mass in their
study assumes the \citet{Kroupa01} IMF, which is about 30\% lower than the
diet Salpeter IMF we adopt here. With the IMF difference corrected, 
our estimate of the dependence of the $z\sim 0$ star formation efficiency 
on halo mass agrees well with their results (see their Table~3).
After correcting for differences in the adopted IMF and halo definitions, 
in the relevant halo mass range our results at $z\sim0$ and $z\sim 1$ also 
agree with those reported by \citet{Heymans06} based on a weak gravitational 
lensing study of the {\it Hubble Space Telescope} GEMS (Galaxy Evolution from 
Morphologies and SED) survey. 
It is worth emphasizing that our results follow from the mass 
distribution of CDM halos and the galaxy assignment required to 
reproduce the observed clustering, while the galaxy lensing results
are from direct measurements of the dark matter halos through their
weak-lensing effects.  The agreement between our results and the lensing
results is therefore, again, impressive.

What determines the halo mass scale where the average conversion efficiency 
reaches a maximum? It could be star formation feedback or preheating. 
\citet{Mo05} argue that present day halos less massive than $10^{12}\hMsun$ 
were embedded in pancakes of $M\sim 5\times10^{12}\hMsun$ at $z\sim 2$, 
whose formation heats and compresses the gas, leading to a cooling time 
longer than the age of the universe at $z\lesssim 2$. Therefore, in halos 
below $10^{12}\hMsun$, there is not much cold gas available for star 
formation, which leads to the drop in the conversion efficiency. Our 
$z\sim 0$ mass scale is roughly consistent with their prediction. On the 
other hand, our results indicate that this halo mass scale shifts to a 
higher mass at $z\sim 1$.  If true, preheating alone may not be sufficient 
to explain this mass scale. The intensity of star formation and the 
subsequent feedback as a function of halo mass and redshift may also be 
an important factor. The shift of the peak of star formation efficiency 
from high-mass halos to low-mass halos with time can be regarded as another 
manifestation of the downsizing pattern seen in star-forming galaxies. 
For both DEEP2 and SDSS galaxies, the drop of conversion efficiency above 
the characteristic halo mass could be in part due to the fact that we only 
consider stars in central galaxies, and in high-mass halos the stellar 
mass contributions from satellite galaxies can be substantial. 
In addition, in high-mass halos gas accretion becomes less efficient
because of the high virial temperature.  We note again that the stellar 
mass results presented here are shown mainly as a proof of concept and any 
conclusions based on these should be regarded as tentative.

\section{Summary and Future Prospects}

We perform HOD modeling of the projected galaxy two-point correlation
functions $w_p(r_p)$ for luminosity threshold samples of DEEP2 and
SDSS galaxies at $z\sim 1$ and $z\sim 0$, respectively.  The HOD
modeling, which converts galaxy pair statistics to relations between
galaxy properties and dark matter halos, reproduces well the galaxy
correlation functions at these two redshifts, including the rise on
small scales seen at $z\sim 1$.

We infer the relationship between central galaxy luminosity and halo mass
at $z\sim 0$ and $z\sim 1$. We find that at both redshifts the mean 
central galaxy luminosity $\meanLc$ increases with halo mass. In low-mass 
halos below $\sim 10^{12}\hMsun$,  $\meanLc$ increases more rapidly with 
halo mass, and the scatter in the central galaxy luminosity in halos of 
fixed mass appears larger, possibly reflecting the broad distribution 
of major star formation epochs in these halos.  We find evolution in the 
HOD in that galaxies at a given halo mass are $\sim$1.4 times more luminous 
at $z\sim 1$ and for a given luminosity the halo mass is $\sim$1.6 times 
greater at $z\sim 0$, with the caution that luminosities of DEEP2 and SDSS 
are in different rest-frame bands. In addition, central $L_*$ galaxies are 
found to be in halos a few times more massive at $z\sim 1$ than at $z\sim 0$.  

For both DEEP2 and SDSS galaxies, there exists a scaling relation between 
$\Mmin$, the characteristic minimum mass of the halo that can host a central 
galaxy above a given luminosity threshold, and $M_1$, the mass scale of 
a halo that on average is able to host one satellite galaxy above the same 
luminosity threshold. There is little difference between the scaling relation 
at $z\sim 1$ ($M_1\simeq 16\Mmin$) and that at $z\sim 0$ ($M_1\simeq 18\Mmin$). 
The fraction of galaxies that are satellites decreases with increasing galaxy 
luminosity at both redshifts. At a fixed luminosity threshold (in units of 
$L_*$) SDSS galaxies have a larger satellite fraction than DEEP2 galaxies, 
e.g., $\sim$20\% versus $\sim$10\% for $L>L_*$ galaxies.

The ultimate goal in modeling galaxy clustering at different redshifts
is to learn about galaxy formation and evolution. These HOD modeling results
provide us with relationships between galaxies and dark matter halos at 
$z\sim 1$ and $z\sim 0$, spanning half the history of the universe.
We use the typical growth of dark matter halos as determined from simulations 
to establish an evolutionary link between DEEP2 and SDSS galaxies and 
extract information about galaxy evolution over the last 7 billion years. 
We establish such a connection, on average, by linking $z\sim 0$ central 
galaxies to $z\sim 1$ central galaxies residing in the progenitor halos of 
the $z\sim 0$ halos. We relate the luminosities of $z\sim 0$ central 
galaxies to those of their $z\sim 1$ progenitors; however, the interpretation 
is complicated somewhat by the different rest-frame selection bands of SDSS 
and DEEP2 galaxies.  As a proof of concept, we use stellar masses determined 
by galaxy color and luminosity to roughly estimate the evolution of the 
dependence of stellar mass on host halo dark matter mass.  At $z\sim 1$ the 
mean fraction of baryons that have been converted into stars is below 
$\sim$15\%, peaking at a halo mass of $10^{12}\hMsun$. At $z\sim 0$ this 
fraction is below $\sim$30\% and the peak of the star formation efficiency 
is at a lower halo mass of $\sim 6\times 10^{11}\hMsun$. We find that, on 
average, the majority of stars in $z\sim 0$ central galaxies in low-mass 
halos (a few $\times 10^{11}\hMsun$) formed between $z\sim 1$ and 
$z\sim 0$, while only a small fraction of stars formed in central galaxies 
of halos as massive as $\sim 10^{13}\hMsun$.  This reflects the 
downsizing pattern seen for star-forming galaxies as a function of 
stellar mass, shown now as a function of their host halo mass. 

The stellar mass results presented here are preliminary, as the galaxy 
samples we model are defined by luminosity, not stellar mass. In addition, 
the SDSS and DEEP2 surveys have different rest-frame selection functions, 
which complicates comparing the results at the two redshifts.
Ideally we would like to have stellar mass-selected galaxy samples at 
different redshifts from surveys that have the same rest-frame selection
functions. While our results apparently agree with other studies in many 
aspects, we regard our conclusions on the details of the 
stellar mass evolution as tentative. Nevertheless, the halo mass-dependent 
evolution trend we infer appears to be robust, and the exercise serves 
as a starting point and guide for future work. 

HOD modeling of galaxy clustering for stellar mass-selected samples
at different redshifts would be a great tool for gaining insight on galaxy
formation and evolution, and we plan to pursue this in future work. 
Unlike the simple estimates presented in this paper, which are based on a 
mean relationship between the final halo mass and progenitor halo mass, 
a comprehensive program would be more sophisticated. Halos at high redshift, 
populated with galaxies as a function of stellar mass in accordance with 
the HOD results, would serve as initial conditions.  Then, assuming no star 
formation, these halos would be evolved forward in time to lower redshift, 
following the merging histories and satellite dynamics determined from 
simulations or analytic models.  This method is similar to the semi-analytic 
galaxy formation model in merging of halos and dynamically evolving satellites,
but here the initial conditions are set by the HOD modeling results. 
This method differs from semi-analytic models in that only dark 
matter evolution and dynamics, not baryon physics, are involved 
in the calculation.  The difference between the evolved HOD and the HOD 
inferred from galaxy clustering at low redshift would then provide a wealth 
of information on star formation as a function of halo mass. For example, the 
contributions to stellar mass in central galaxies can be separated --- what 
fractions are from star formation, from central galaxies in other halos that 
merge with them, and from minor mergers with satellite galaxies.
We would also infer the average star formation history in satellite galaxies. 
If galaxy clustering is modeled at a series of redshifts, then the 
average {\it continuous} star formation history and the average stellar 
mass growth history from other modes (i.e., major and minor
mergers) as a function of halo mass can be determined. Such
empirically-derived results would shed light on the physics of galaxy 
formation, and comparisons with predictions of galaxy formation models
would provide stringent tests for galaxy formation theory.
An example of the effort starting along some of these lines is
\citet{White07}, which uses $N$-body simulations in combination with 
HOD modeling of clustering to put constraints on merging of luminous 
galaxies from $z\sim 0.9$ to $z\sim 0.5$.

Additionally, the envisioned program may also help test the role of 
environment on galaxy formation and evolution. One of the main assumptions 
in the current HOD framework is that the galaxy content in halos depends 
only on the halo mass and is statistically independent of the halo's larger 
scale environment. \citet{Gao05} find in simulations a signature of 
age-dependent halo clustering (the so-called halo assembly bias, see 
also \citealt{Croton07,Jing07,Harker06,Wechsler06,Wetzel07,Zhu06,Gao07}). 
In general, the environmental effect on halo clustering is essentially
negligible for halos above the nonlinear mass scale and becomes stronger for 
lower mass halos. With the method outlined above we may be able to test the 
effect of large-scale environment on galaxy evolution. It is likely that the 
environment-dependent halo clustering does not play a significant role in 
interpreting the observed galaxy clustering at high redshift due to the fast 
drop of the nonlinear mass (e.g., from $z=1$ to $z=0$, the nonlinear 
mass increases by a factor of $\sim$25, while halos only grow by a factor 
of about 2 in a large range of mass). When we passively evolve a high-redshift 
HOD, which is insensitive to the environment, to lower redshift,
the environmental effect during the evolution can be fully taken into account.
Comparing then the evolved HOD (which accounts for the environmental effect)
with the HOD inferred from the low redshift data (which ignores the
environmental effect) may provide insight on how environment shapes the 
HOD and determines properties of galaxies in low-mass halos, increasing
the constraining power of galaxy clustering on galaxy formation processes.

\acknowledgments 

We are grateful to Christopher Willmer and Kevin Bundy for providing stellar 
mass estimates of DEEP2 galaxies. We thank the referee David Weinberg for 
many useful discussions and comments that improved the paper.
We thank Charlie Conroy for useful comments on an earlier draft of this 
paper and Michael Brown, Jeffrey Newman, Jeremy Tinker, and Martin White for 
helpful conversations.

Z. Z. and A. L. C. are supported by NASA through Hubble
Fellowship grants HF-01181.01-A and HF-01182.01-A, respectively, awarded
by the Space Telescope Science Institute, which is operated by the
Association of Universities for Research in Astronomy, Inc., for NASA
under contract NAS 5-26555. 

DEIMOS was funded by a grant from CARA (Keck
Observatory), an NSF Facilities and Infrastructure grant (AST92-2540),
the Center for Particle Astrophysics, and by gifts from Sun
Microsystems and the Quantum Corporation. 
The DEEP2 data presented herein were obtained at the W. M. Keck Observatory, 
which is operated as a scientific partnership among the California Institute of
Technology, the University of California, and the National Aeronautics
and Space Administration. The Observatory was made possible by the
generous financial support of the W.M. Keck Foundation. The DEEP2 team
and Keck Observatory acknowledge the very significant cultural role
and reverence that the summit of Mauna Kea has always had within the
indigenous Hawaiian community and appreciate the opportunity to
conduct observations from this mountain.

Funding for the SDSS and SDSS-II has been provided by the Alfred P. Sloan 
Foundation, the Participating Institutions, the National Science Foundation, 
the US Department of Energy, the National Aeronautics and Space 
Administration, the Japanese Monbukagakusho, the Max-Planck Society, and 
the Higher Education Funding Council for England. The SDSS Web site is 
http://www.sdss.org/.

The SDSS is managed by the Astrophysical Research Consortium for the 
Participating Institutions. The Participating Institutions are the 
American Museum of Natural History, Astrophysical Institute Potsdam, 
University of Basel, Cambridge University, Case Western Reserve 
University, University of Chicago, Drexel University, Fermilab, 
the Institute for Advanced Study, the Japan Participation Group, 
Johns Hopkins University, the Joint Institute for Nuclear Astrophysics, 
the Kavli Institute for Particle Astrophysics and Cosmology, 
the Korean Scientist Group, the Chinese Academy of Sciences (LAMOST), 
Los Alamos National Laboratory, the Max-Planck-Institute for Astronomy (MPIA), 
the Max-Planck-Institute for Astrophysics (MPA), New Mexico State University, 
Ohio State University, University of Pittsburgh, University of Portsmouth, 
Princeton University, the United States Naval Observatory, and the 
University of Washington.

\appendix

\section{Bias related to the assumption of ``one galaxy per halo''}

\begin{figure}[t]
\epsscale{1.0}
\plotone{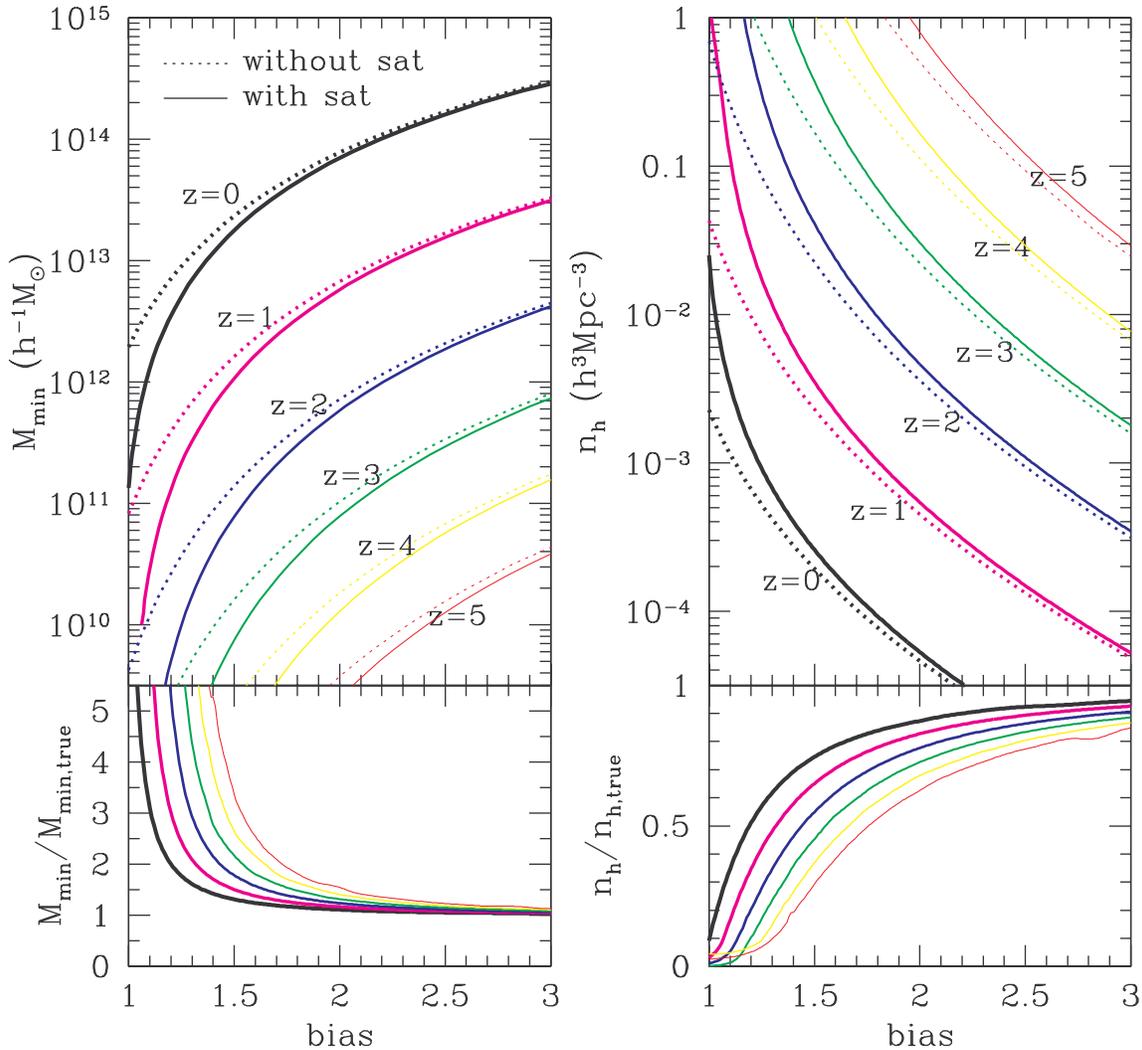}
\epsscale{1.0}
\caption[]{
\label{fig:appendix}
{\it Left:} Comparison between the minimum mass $\Mmin$ of the host
halos under the assumption of one galaxy per halo (i.e., no 
satellite galaxies) and the true value at a measured galaxy bias factor
for galaxy samples at different redshifts. The bottom panel shows the
ratio of the $\Mmin$ from one galaxy per halo to the true value. 
{\it Right:} Same as the left panels, but for comparisons between 
the inferred number densities of halos (see text for details).
}
\end{figure}

Given a measured two-point correlation function for a galaxy population, 
HOD modeling is the ideal path to infer information about the host halos. 
However, this involves non-trivial modeling efforts, and therefore 
approximations are often used in the literature.  One commonly-used method 
assumes that there is one galaxy per halo and connects galaxies to halos 
through the large-scale bias factor. This assumes that all galaxies in the 
sample are central galaxies, while in reality a small fraction are satellite 
galaxies. Using a simple HOD model, we quantify the resulting systematic 
bias in the inferred halo mass and the duty cycle of galaxies from this 
assumption.

We consider galaxy samples defined by a lower threshold in luminosity or
stellar mass. The usual procedure associated with the ``one galaxy per 
halo'' assumption is as follows.  The large-scale bias factor of galaxies
is obtained from the ratio of the measured galaxy two-point correlation 
function and the theoretical matter two-point correlation function under 
the assumed cosmology. This large-scale bias factor is the same as that of 
halos above a mass threshold. Since for an assumed cosmology the latter 
is a known function of mass threshold 
(e.g., \citealt{Mo96,Sheth01,Tinker05}), the halo mass 
threshold $\Mmin$ can be determined. The number density of halos above
the mass threshold $\Mmin$ can be inferred from the halo mass function
(e.g., \citealt{Sheth99,Jenkins01}). The ratio of the observed galaxy number 
density to the inferred halo number density is used to determine the 
duty cycle of galaxies. More exactly, the ``one galaxy per halo'' assumption
should be called ``$f$ galaxy per halo,'' where $f$ is a positive constant 
smaller than unity representing the duty cycle. 

If satellite galaxies are taken into account, using the same galaxy bias 
factor measured from the data, a different $\Mmin$ would be inferred.
To see how large the difference is, we assume a simple mean occupation 
function to represent the ``truth;'' this model is the sum of a step 
function for central galaxies and a power law for satellite galaxies, 
$\NavgM=1+M/M_1$ for $M>\Mmin$. Following galaxy formation model 
predictions (e.g., \citealt{Kravtsov04,Zheng05}) and HOD modeling results 
(e.g., this paper; \citealt{Zehavi05}), we use $M_1=20\Mmin$. The above 
$\Navg$ can have an overall normalization factor smaller than unity (duty 
cycle factor), and none of our results below change.

The bias factor $b$ for a given galaxy sample (in which all galaxies reside 
in halos of mass $M>\Mmin$) can be calculated as
\begin{equation}
\label{eqn:bias}
b(>\Mmin)=\left.
          \int_{\Mmin}^{+\infty} dM \frac{dn}{dM} \NavgM b_h(M) 
          \right/ 
          \int_{\Mmin}^{+\infty} dM \frac{dn}{dM} \NavgM,
\end{equation}
where $dn/dM$ is the halo mass function and $b_h(M)$ is the 
bias factor of halos of mass $M$ (following \citealt{Sheth99}). 
Solid lines in the top left panel of Figure~\ref{fig:appendix} show the 
relation between $\Mmin$ and $b(>\Mmin)$ at different redshifts. 
Assuming one galaxy per halo, the bias is calculated by replacing $\NavgM$ 
with unity in equation~(\ref{eqn:bias}). Dotted lines in the top left 
panel of Figure~\ref{fig:appendix} show this relation. At a 
given bias factor (measured from the galaxy two-point correlation function), 
the value of $\Mmin$ with the assumption of no satellite galaxies ({\it dotted 
lines}) is always higher than the ``true'' value ({\it solid lines}) because 
massive halos (with high halo bias) are weighted more as the number of 
satellites increases with halo mass. The bottom left panel of 
Figure~\ref{fig:appendix} shows the ratio of the $\Mmin$ from the ``one 
galaxy per halo'' method to the true value at each redshift, as a function 
of the galaxy bias. When the bias is high, the corresponding halos are in 
the exponential tail of the mass function and the weight from the satellite 
galaxies residing in massive halos is less significant, therefore, the 
value of $\Mmin$ inferred from the approximate method approaches its true 
value. However, the difference increases for lower galaxy biases, 
corresponding to the shallower part of the halo mass function, and at higher
redshift. For instance, at $z=0$, for a galaxy sample with $b=1.2$, the 
estimated $\Mmin$ can be a factor of 2 higher than the true value.

The difference in the inferred values of $\Mmin$ leads to a difference in
the inferred number densities of halos above $\Mmin$. The halo number 
density $n_h(>\Mmin)$ is calculated as 
\begin{equation}
\label{eqn:numden}
n_h(>\Mmin)= \int_{\Mmin}^{+\infty} dM \frac{dn}{dM}.
\end{equation}
The right panels of Figure~\ref{fig:appendix} compare the inferred (from 
``$f$ galaxy per halo'') and the true number densities of halos as a 
function of galaxy bias. For a given galaxy bias, the halo number density 
derived from the approximate $\Mmin$ is always lower than the true value 
because without satellites in massive halos, a larger value of $\Mmin$ is 
needed to reach the given bias factor. As with $\Mmin$, the estimated halo 
number density approaches the true value for highly biased galaxy samples, 
while for low bias factors, the difference can again be substantial. For 
example, at $z=0$, the number density of host halos for galaxies with 
$b=1.2$ is underestimated by a factor of 2. As a consequence, the duty 
cycle (the fraction of halos hosting galaxies of that type) is overestimated 
by using the underestimated halo number density.

For galaxy samples defined by a lower threshold (e.g. in luminosity),
Figure~\ref{fig:appendix} can be used to better estimate the minimum host 
halo mass and number density from the measured galaxy bias value, instead of
assuming one galaxy per halo.

{}

\end{document}